\documentclass[aps,twocolumn,showpacs,superscriptaddress,groupedaddress]{revtex4} 

\usepackage{soul}
\usepackage{amsmath}
\usepackage{amssymb}
\usepackage{graphicx}
\usepackage{morefloats}
\usepackage{soul}
\usepackage[usenames,dvipsnames]{xcolor}
\usepackage{blkarray}
\usepackage{verbatim}
\usepackage{hyperref}
\usepackage{subfigure}
\usepackage{natbib}
\usepackage{wrapfig}

\hypersetup{colorlinks=true, citecolor=orange, urlcolor=cyan, linkcolor=blue}
\newcommand{\NL}{N_\text{L}}

\makeatletter
\def\Dated@name{ }
\makeatother
\begin{document}

\title{%Nested mess: t
%Thermodynamics disentangles conflicting notions of nestedness in ecological networks 
%\textcolor{blue}{A nested mess tale
%: on the perils of measuring the nestedness %of a system}
The ambiguity of nestedness under soft and hard constraints

}

\author{M. Bruno*}
% \email{matteo.bruno@imtlucca.it}
\affiliation{IMT School for Advanced Studies, P.zza S. Francesco 19, 55100 Lucca (Italy)}
\author{F. Saracco}
\affiliation{IMT School for Advanced Studies, P.zza S. Francesco 19, 55100 Lucca (Italy)}
\author{D. Garlaschelli}
\affiliation{IMT School for Advanced Studies, P.zza S. Francesco 19, 55100 Lucca (Italy)}
\affiliation{Lorentz Institute for Theoretical Physics, University of Leiden, Niels Bohrweg 2, 2333 CA Leiden (The Netherlands)}
\author{C. J. Tessone}
\affiliation{URPP Social Networks, University of Z\"urich, Andreasstrasse 15, CH-8050 Z\"urich (Switzerland)}
\author{G. Caldarelli}
\affiliation{IMT School for Advanced Studies, P.zza S. Francesco 19, 55100 Lucca (Italy)}
\affiliation{European Centre for Living Technology, Universit\`a di Venezia “Ca' Foscari”, S. Marco 2940, 30124 Venice (Italy)}
\affiliation{Catchy srl, Talent Garden Poste Italiane, Via Giuseppe Andreoli 9, 00195 Rome (Italy)}
\affiliation{Istituto dei Sistemi Complessi CNR, Dip. Fisica, Universit\`a Sapienza, P.le Aldo Moro 2, 00185 Rome (Italy)}
\date{* email: \textit{matteo.bruno@imtlucca.it}}

\begin{abstract}
Many real networks feature the property of nestedness, i.e. the neighbours of nodes with a few connections are hierarchically nested within the neighbours of nodes with more connections. Despite the abstract simplicity of this notion, various mathematical definitions of nestedness have been proposed, sometimes giving contrasting results. Moreover, there is an ongoing debate on the statistical significance of nestedness, since random networks where the number of connections (degree) of each node is fixed to its empirical value are typically as nested as real %-world
ones. By using only ergodic and unbiased null models, we %explore how this kind of approach can lead to wrong conclusions if not performed correctly. 
propose a clarification that exploits the recent finding that random networks where the degrees are enforced as hard constraints (microcanonical ensembles) are thermodynamically different from random networks where the degrees are enforced as soft constraints (canonical ensembles).
%We show that if the real network is perfectly nested, then the two ensembles are trivially equivalent and the observed nestedness, independently of its definition, is indeed an unavoidable consequence of the empirical degrees. 
Indeed, alternative definitions of nestedness can be  negatively correlated in the microcanonical one, while being positively correlated in the canonical one.
This result disentangles distinct notions of nestedness captured by different metrics and highlights the importance of making a principled choice between hard and soft constraints in null models of ecological networks.
\end{abstract}

\maketitle

\section{Introduction}\label{sec:Intro}
Network theory provides a simplified representation of a variety of complex systems, i.e. systems composed by many elements whose mutual interactions create new and emergent behaviours. The network description, despite its simplification, allows to detect and measure collective patterns, independently of the nature of the underlying interactions~\cite{Boccaletti2006, Caldarelli2007, Newman2010,SquartiniTiziano2017}.

Amongst the quantities analysed in network theory, nestedness~\cite{MARIANI20191} is one of the most elusive. 
It was originally observed in biogeography~\cite{Hulten1937,Patterson1986,Atmar1993} where less frequently observed species are assumed to occupy a niche of the habitats occupied by more ubiquitous species. 
In terms of the resulting ecological network, nestedness is loosely defined as the observation that the neighbours of nodes with a few connections (lower degree) are typically a subset of the neighbours of nodes with more connections (higher degree).
Generalized as such, nestedness has been detected in other networks as well, e.g. in trade networks~\cite{Tacchella2012, Konig2014, Saracco2016}, interbank networks~\cite{Konig2014,Soramaki2007}, social-media information networks~\cite{Borge-Holthoefer2017}, and mutualistic ecological networks~\cite{Bascompte2003, MARIANI20191}. In previous works~\cite{Bastolla2009, rezende2007}, nestedness has been found to be highly correlated with the stability of the ecosystem under different types of disturbances and perturbations.
The ubiquity and structural importance of nestedness naturally raises some fundamental questions regarding the possible mechanism generating nested patterns in real networks~\cite{Suweis2013, Gracia-Lazaro2018}. 
Actually, while the intuitive notion of nestedness is straightforward, its mathematical definition is not trivial and different metrics, focusing on different aspects, have been proposed. One of the most popular metrics is NODF (\emph{Nestedness measure based on Overlap and Decreasing Fill},~\cite{Almeida-Neto2008a}), which considers the (normalized) overlap between pairs of nodes in the same layer of a bipartite network. Such a definition was later adjusted in order to increase its robustness~\cite{Bastolla2009,MARIANI20191}.

An alternative measure has been proposed by looking at certain spectral properties of the adjacency matrix of a bipartite network. Since it can be shown that, when the degree sequence is constrained on one of the two layers of the network, the spectral radius is maximum for the perfectly nested network~\cite{Bhattacharya2008b}, the spectral radius itself was proposed as a measure of nestedness \cite{Staniczenko2013} (in the following SNES, i.e. Spectral NEStedness).

Beside the quest for  measures properly capturing the sense of nestedness, some researchers focus on disentangling the role of other network properties from the nestedness itself. In this sense, some early contributions focused on the comparison of the measurements with some null models, i.e. statistical models that display some properties of the real system, in order to have a tailored benchmark. Null models were used, for instance, to detect the effect of the degree sequence~\cite{Ulrich2007, Ulrich2009,Ulrich2012,Gotelli2012}. Actually, to properly define a null model, the approaches to follow can be, essentially, 2: 1) to impose constraints \emph{exactly} or 2) to impose constraints \emph{on average}, respectively microcanonically and canonically, according to the Statistical Physics jargon. Theoretical tools from statistical physics are not new for the analysis of ecological system: they are commonly used to investigate patterns in biological networks, targeting, from time to time, hierarchical systems~\cite{Agliari2015}, bipartite structures~\cite{Sollich2014} and topological properties of scale-free networks~\cite{Agliari2011}.\\. Regarding the former case, beside various approaches, the algorithm of Ref.~\cite{Strona2014} only was shown to be ergodic~\cite{Carstens2015}, i.e. to visit uniformly the phase space and, thus, to provide unbiased predictions. 
Instead, in the \emph{canonical} approach, constraints are satisfied \emph{on average} and, due to its derivation, is ergodic \emph{a priori} ~\cite{SquartiniTiziano2017,Cimini2018}.  
The canonical approach allows for some noise in the data: indeed if there is some noise, an existing pollinator-plant interaction may not be detected and the microcanonical approach will not consider the real configuration among the possible ones, while the canonical one will. Using the canonical approach, 
Ref.~\cite{Jonhson2013} compared the metric introduced in Ref.~\cite{Bastolla2009} with a null model preserving the degree sequence and found that in most of the cases the degree sequence is responsible for the high value of the nestedness (actually, the null model implemented in Ref.~\cite{Jonhson2013} is out the regime of validity). The recent contribution of Payrató-Borràs et al.~\cite{payrato2019breaking} came to similar conclusions, using an improved null model still preserving the degree sequence, but valid for any level of link density of the network, although using an approximated formula for the average of NODF in the ensembles; subsequently the same group enlarged their analysis to a wider number of nestedness metrics~\cite{payratoborras2020measuring}.\\

As presented above, in the literature, several papers  compared the measurements with various null models~\cite{Ulrich2007, Ulrich2009,Ulrich2012,Gotelli2012, Jonhson2013, Strona2018,payrato2019breaking,payratoborras2020measuring}, but rarely the issue of ergodicity was targeted. For the first time here, we investigate the differences of the micro- and canonical approach in discounting the degree sequence for the analysis of the nestedness.
Let us remark that both the null models implemented are \emph{ergodic}, i.e. they explore the phase space homogeneously, in order to have unbiased and unequivocal results.\\

In the present paper, we provide an example in which two null models ergodically discounting the same information, i.e. the degree sequence, display opposite correlations between two among the most used nestedness measures. Beside providing another example of the statistical ensemble inequivalence, the main message of our manuscript is that choosing to quantify the amount of nestedness is a subtle task that has to be carried out carefully. Indeed, only being aware of the behaviour and the peculiar properties of the various approaches and options permits to derive the right conclusions from the analyses: our paper provides the necessary knowledge to handle properly the study of the nestedness of a real system. In this sense, we do not provide any univocal indication on which nestedness definition measure should be used or on which is the correct way to discount the information encoded in the degree sequence: both the definitions analysed have their own sense and both the null models examined satisfy their own rationale. Nevertheless, it is crucial to know the properties of the various tools that one is handling in order to derive the proper conclusions from the nestedness analysis of a real system.

%These findings indicate that a bigger effort must be done if null models need to be used to assess the statistical significance of the nestedness of a system, and their use must be justified thoroughly.} 

\section{Methods}\label{sec:Methods}
A bipartite network is defined by two sets of nodes $\text{L}$ (of size $N_\text{L}$) and $\Gamma$ (of size $N_\Gamma$) called \emph{layers} and by the prescription that connections are allowed only between the layers and not inside them.  
Thus, a bipartite network can be univocally described by its biadjacency matrix $\textbf{M}$, i.e. an $(N_\text{L}\times N_\Gamma)$-matrix, whose entries $m_{i\alpha}=1$ if a link exists between $i\in N_\text{L}$ and $\alpha\in\Gamma$ and $m_{i\alpha}=0$ otherwise. We will call a network perfectly nested (PNN, \emph{Perfectly Nested Network}) if for every pair of nodes $i$, $j$ belonging to the same layer with degrees $d_i$, $d_j$, if $d_i \leq d_j$ then all neighbours of $i$ are also neighbours of $j$. This type of network is also called \textit{chain graph} \cite{Bhattacharya2008b} or \textit{double nested graph} \cite{bell2008graphs}. %Aggiungere citazione
In the following we shall use the previous definitions for generic biadjacency matrices and related quantities, but we shall add an asterisk $*$ whenever considering quantities measured on real networks. 

\subsection{Nestedness measures}\label{subsec:nestedness_measures}

\subsubsection{NODF}
One of the most popular measure of nestedness, namely the \emph{Nestedness as a measure of Overlap and Decreasing Fill} (NODF) was introduced in 2008 by Almeida-Neto et al. \cite{Almeida-Neto2008a}. Such measure is based on the overlap between the neighbourhoods of nodes with different degrees. Given a generic bipartite graph $G_\text{Bi}$, the NODF expression reads 
% \begin{equation}\label{eq:NODF}
% \operatorname{NODF}(\mathbf{M}) = \frac{1}{K}\Bigg[\sum\limits_{i,j=1}^{N_\text{L}}\Bigg(\theta(k_i - k_j)\cdot \frac{\sum\limits_{\alpha=1}^{N_\Gamma} m_{i\alpha}m_{j\alpha}}{k_j}\Bigg)
% + \sum\limits_{\alpha,\beta=1}^{N_\Gamma}\Bigg(\theta(h_\alpha - h_\beta) \cdot \frac{\sum\limits_{i=1}^{N_\text{L}} m_{i \alpha}m_{i \beta}}{h_\beta}\Bigg)\Bigg]
% \end{equation}
\begin{equation}\label{eq:NODF}
\begin{split}
    \operatorname{NODF}(\mathbf{M}) &= \frac{1}{K}\Bigg[\sum\limits_{i,j=1}^{N_\text{L}}\Bigg(\theta(k_i - k_j)\cdot \frac{\sum\limits_{\alpha=1}^{N_\Gamma} m_{i\alpha}m_{j\alpha}}{k_j}\Bigg)\\ 
    &+ \sum\limits_{\alpha,\beta=1}^{N_\Gamma}\Bigg(\theta(h_\alpha - h_\beta) \cdot \frac{\sum\limits_{i=1}^{N_\text{L}} m_{i \alpha}m_{i \beta}}{h_\beta}\Bigg)\Bigg]
\end{split}
\end{equation}
where $K = \frac{N_\text{L}(N_\text{L}-1) + N_\Gamma(N_\Gamma-1)}{2}$ is a normalization factor to let the measure go from $0$ to $1$, $k_i$ and $h_\alpha$ are respectively the degrees of node $i\in\text{L}$ and $\alpha\in\Gamma$, and $\theta$ is the Heaviside step function with the convention $\theta(0) = 0$. The step function ensures that the overlap is only counted when the degrees of the nodes are different and that the denominator is the minimum of the two vertices' degrees.

\subsubsection{Stable-NODF}
Due to the instability of the previous measure, with respect to small fluctuations on the degrees of the nodes, another version was proposed in \cite{MARIANI20191}. The difference relies in considering also the contributions coming from couples of nodes with equal degrees; we will call it stable-NODF or sNODF. It is calculated as

% \begin{equation}\label{eq:sNODF}
% \operatorname{sNODF}(\mathbf{M}) = \frac{1}{K}\Bigg[\sum\limits_{i<j}^{\NL}\Bigg(\frac{\sum\limits_{\alpha=1}^{N_\Gamma} m_{i\alpha}m_{j\alpha}}{\min(k_i,k_j)}\Bigg)
% + \sum\limits_{\alpha<\beta}^{N_\Gamma}\Bigg( \frac{\sum\limits_{i=1}^{\NL} m_{i \alpha}m_{i \beta}}{\min(h_\alpha, h_\beta)}\Bigg)\Bigg]
% \end{equation}
\begin{equation}\label{eq:sNODF}
\begin{split}
    \operatorname{sNODF}(\mathbf{M}) &= \frac{1}{K}\Bigg[\sum\limits_{i<j}^{\NL}\Bigg(\frac{\sum\limits_{\alpha=1}^{N_\Gamma} m_{i\alpha}m_{j\alpha}}{\min(k_i,k_j)}\Bigg)\\
    &+ \sum\limits_{\alpha<\beta}^{N_\Gamma}\Bigg( \frac{\sum\limits_{i=1}^{\NL} m_{i \alpha}m_{i \beta}}{\min(h_\alpha, h_\beta)}\Bigg)\Bigg] \: .
\end{split}
\end{equation}
where $K$ is the same normalization factor as in \ref{eq:NODF} and the denominator this time is the minimum between the two degrees, that in \ref{eq:NODF} was guaranteed by the theta step function.

\subsubsection{Spectral nestedness} A recently proposed measure of nestedness \cite{Staniczenko2013} considers the spectral radius of the network, i.e. the largest eigenvalue $\lambda$ of the adjacency matrix. We will thus call it spectral nestedness (SNES). The adjacency matrix $\mathbf{A}$ of a bipartite network can be expressed in terms of the biadjacency matrix as
\unexpanded{
\begin{equation*}
\mathbf{A}=
\left(
    \begin{array}{c|c}
    \mathbf{0}_{N_\text{L}\times N_\text{L}} & \mathbf{M}\\
    \hline
    \mathbf{M}^T & \mathbf{0}_{N_\Gamma\times N_\Gamma}
    \end{array}
\right),
\end{equation*}
}
where $\mathbf{M}^T$ is the transpose of the biadjacency matrix $\mathbf{M}$ and $\mathbf{0}_{N\times N}$ is a $N\times N$-matrix whose elements are all zeros. Note that the adjacency matrix of the network is symmetric, yielding all real eigenvalues. The definition is based on two main theoretical results:
\begin{itemize}
    \item The bipartite network that has the maximum eigenvalue in the set of connected networks with given $n$ nodes and $L$ links is a perfectly nested network \cite{bell2008graphs};
    \item Among all bipartite networks with a given degree sequence on one of the two layers, the one that maximizes the spectral radius is the PNN, defined at the beginning of the present section%perfectly nested one
    ~\cite{Bhattacharya2008b}.
\end{itemize}

\subsubsection{Normalized spectral nestedness}
The spectral radius, though, has a strong dependence on the size of the network and on its density. It is well known that the maximum eigenvalue of a bipartite network with $L$ links is bounded from above by $\sqrt{L}$ and that the only network for which $\lambda(\mathbf{M}) = \sqrt{L(\mathbf{M})}$ (if it exists) is a complete bipartite network~\cite{bell2008graphs,Bhattacharya2008b}.

For this reason we decide to introduce nSNES,
where we normalize the measure with the square root of the number of edges:
\begin{equation}\label{eq:SNES}
    \operatorname{nSNES}(\mathbf{M})=\frac{\operatorname{SNES}(\mathbf{M})}{\sqrt{L(\mathbf{M})}} = \frac{\lambda(\mathbf{M})}{\sqrt{L(\mathbf{M})}} \: .
\end{equation}

Although the nSNES ranges from 0 to 1, the drawback of this normalization is that a perfectly nested matrix that is not full will not have a perfect score of 1.

\subsection{Null models}\label{subsec:null_models}
In the present paper, we aim at understanding the role of the degree sequence in the formation of bipartite nested structures.
Thus, we would need a sort of network benchmark with the same degree sequence, but otherwise maximally random. This approach has strong similarities with Statistical Mechanics: actually, the recipe is to build an \emph{ensemble} and fix the node degrees on it. As in the standard Statistical Mechanics, those constraints can be imposed on average, as in the canonical construction~\cite{Park2004,Garlaschelli2008, Squartini2011,SquartiniTiziano2017, Cimini2018}, or considering stricter constraints, as in the microcanonical formulation~\cite{Strona2014,Carstens2015}. The two approaches are known to be non equivalent~\cite{DenHollander2000,Barre2007,Campa2009,Radin2013,Touchette2015,Squartini2015c, Squartini2015a} and indeed such non equivalence is going to be crucial in the following. 

After the randomization with the null models, our aim is to quantify the statistical significance of the measures by computing the z-scores of the measures, calculated as
\begin{equation*}
    z(X)=\dfrac{X-\langle X \rangle}{\sigma_X}
\end{equation*}
where $\sigma_X$ is the standard deviation and X the considered quantity.

\subsubsection{The canonical approach: the Bipartite Configuration Model}
The Bipartite Configuration Model (\textit{BiCM}~\cite{Saracco2015}) is the bipartite extension of the entropy based null model
\cite{Park2004,Garlaschelli2008, Squartini2011,SquartiniTiziano2017, Cimini2018}. The strategy is inspired by work by Jaynes~\cite{Jaynes1957}, which derived the canonical ensemble of Statistical Mechanics from Information Theory principles. The recipe is pretty simple: first, define an ensemble of all possible physical configurations, and then maximize its Shannon entropy constraining the relevant information about the system (in the case of Information Theory, the energy): the result is exactly the canonical ensemble. 
The maximization of the Shannon entropy represents the crucial step: it can be interpreted as assuming maximal ignorance about the the non constrained degrees of freedom of the system.\\

Following the same strategy, starting from a real network, we can define $\mathcal{M}$ the ensemble of all possible biadjacency matrices with the same number of nodes (nodes represent the volume in Statistical Mechanics). The Shannon entropy associated to the ensemble is $S=-\sum_{\textbf{M}\in \mathcal{M}} P(\textbf{M})\ln P(\textbf{M})$ and we can maximize it, constraining the degree sequence (Note that while in the Jaynes derivation of the Statistical Mechanics, the constraints, i.e. the energy, was a global one, the degree sequence represents a local one. Actually, the local constraint is responsible for the nonequivalence of the microcanonical and canonical ensembles.). The entropy maximization leads to an exponential probability for a generic biadjacency matrix $\textbf{M}$: \begin{equation}\label{eq:p_bicm}
P(\mathbf{M})=\frac{e^{-H(\vec{\theta},\:\vec{C}(\mathbf{M})})}{Z(\vec{\theta})},
\end{equation}
where $\vec{C}(\mathbf{M})$ is the vector of constraints and $\vec{\theta}$ the associated Lagrangian multipliers~\cite{Park2004}. At this level the formula \ref{eq:p_bicm} is just formal, in the sense that the value of the Lagrangian multipliers is unknown. At the end of the day, we want a ``tailored" benchmark for our real network, i.e. something with the same degree sequence, but otherwise completely random. In this sense, it is natural to maximize the likelihood of the real network in order to get the value of $\vec{\theta}$~\cite{Garlaschelli2008,Squartini2011}. If $\vec{C}(\textbf{M}^*)$ is the value of $\vec{C}$ measured on the real network, the previous condition is equivalent to impose $\langle \vec{C}(\vec{\theta})\rangle=\sum_{\mathbf{M}\in\mathcal{M}}P(\textbf{M})\vec{C}(\textbf{M})=\vec{C}(\textbf{M}^*)$.\\ 
%Interestingly enough, constraints linear in the adjacency matrix (as the degree sequence)  
The exact solution for the probability $P(\textbf{M})$ can be factorised as the product of probabilities per possible link:
\begin{equation}\label{eq:factorization}
P(\mathbf{M})=\prod_{i, \alpha}p_{i\alpha}^{m_{i\alpha}}(1-p_{i\alpha})^{1-m_{i\alpha}},
\end{equation}
where $p_{i\alpha}$ is the probability of existence of the link connecting nodes $i$ and $\alpha$. Let us remark that the factorisation~(\ref{eq:factorization}) is possible only when the constraints are linear in the biadjacency matrix. For other nonlinear contraints, the probability per link may not be analytical and other methods are necessary to obtain the probability per graph (see for instance~\cite{Fischer2015}).

In the case of the BiCM, $p_{i\alpha}$ is a function of $x_i$ and $y_\alpha$, which are simple reparametrizations of the Lagrange multipliers associated to the observed degrees ($k_i$ and $h_\alpha$ respectively): 
\begin{equation}
p_{i\alpha}=\frac{x_iy_\alpha}{1+x_iy_\alpha}.
\label{prob}
\end{equation}
Their numerical value is determined by solving the likelihood-maximization equations:

\begin{equation}\label{sys}
\left\{ 
\begin{array}{ll}
\langle k_i\rangle&=\sum_\alpha p_{i\alpha} = k_i^*,\:\: i=1\dots \NL\\
&\\
\langle h_\alpha\rangle&=\sum_i p_{i\alpha} = h_\alpha^*,\:\: \alpha=1\dots N_\Gamma
\end{array}
\right.,
\end{equation}
$k_i$ and $h_\alpha$ being the degree of the node $i$ and $\alpha$ respectively.

\subsubsection{The microcanonical approach: the Curveball algorithm}

The microcanonical approach, differently from the BiCM, keeps the degrees of all nodes in the system constant. In a sense, it has a stricter ensemble (just all configurations with the given degree sequence are allowed) and all allowed configurations have the same probability. Such approach is computationally costly since the probabilities of links in the system are not pairwise independent and the fastest way of spanning the ensemble of networks with a given degree sequence relies on swapping endpoints of links iteratively. In the present manuscript, the ensemble was sampled using the strategy of \cite{Strona2014}.

We will refer to this model as \textit{Curveball}, as in the original paper; in~\cite{Carstens2015} it was shown that such approach is ergodic.

\section{Results}\label{sec:Results}
In this section we are going to present the results of our analyses on artificial and real networks.
To test the measures and models, we analyze a set of 40 pollination networks taken from the \textit{Web of Life} dataset (\label{weboflife}\href{www.web-of-life.es}{www.web-of-life.es}). They represent ecological mutualistic networks of plant-pollinators. 
From the set of all available pollination networks in \textit{Web of Life}, we selected only the binary ones, in order to avoid issues regarding binarisation.
All of the considered networks %are binary, and they 
are generally of small size, the smallest being of only 20 nodes while the biggest one consists of 1500 nodes. The density of the networks varies between 0.01 and 0.5.
%Interestingly enough
For the sake of completeness we remark that only 24 out of 40 networks of our dataset are actually made of a single connected component, the other including few disconnected components with more than one node.
In the following, we compare the various measures and state their significance respect to the various null models.

\begin{figure*}[ht!]
    \centering
    \includegraphics[width=\textwidth]{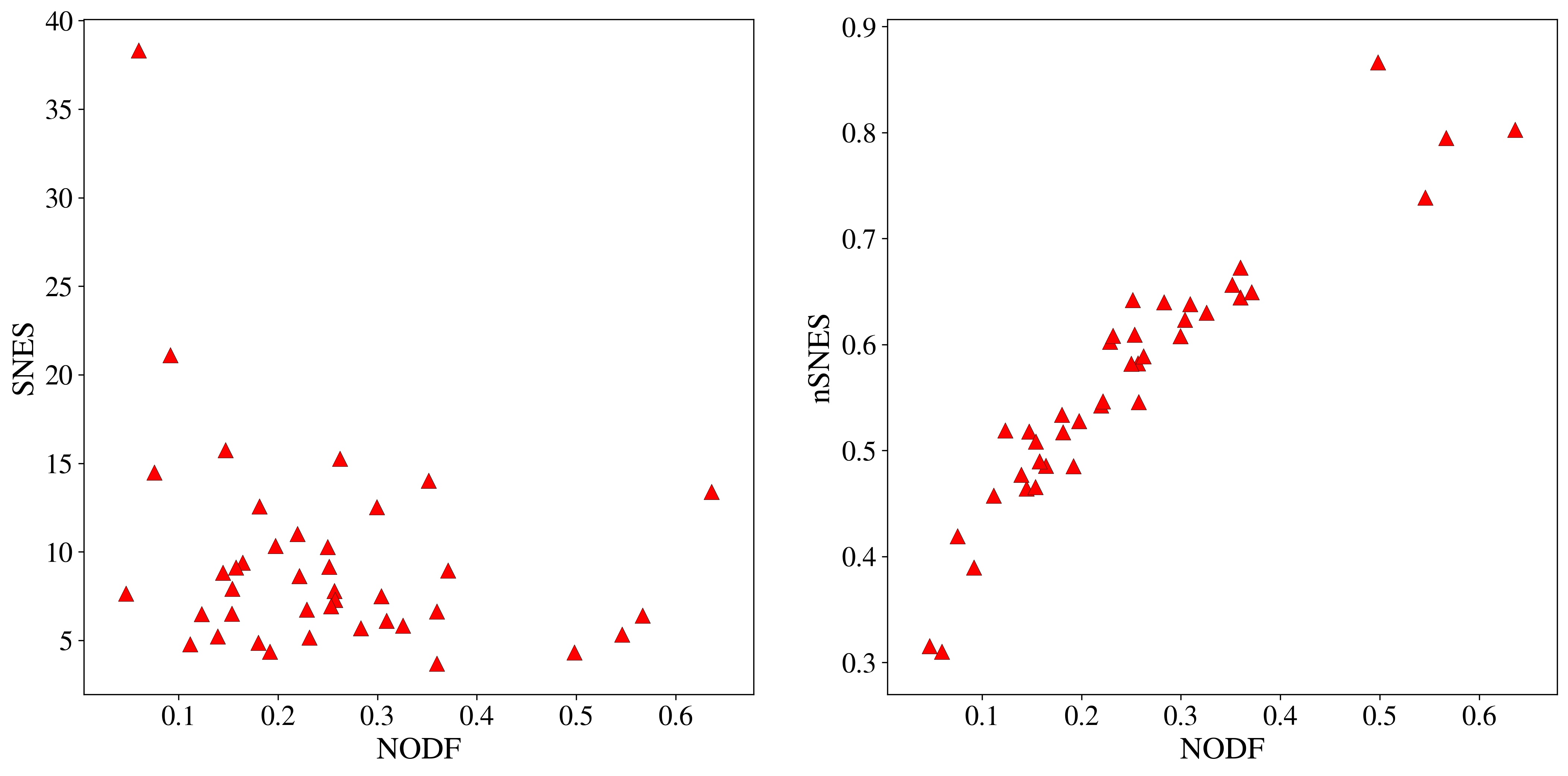}
    \caption{NODF vs SNES (left) and vs nSNES (right) for the 40 networks of the Bascompte dataset. Spearman correlation coefficients are, respectively, -0.23 and 0.96. In fact similar relations hold when considering sNODF instead of NODF: evidences can be found in the Supplementary Fig. A.}% \ref{fig:bascompte_sNODF_vs_SNES}.}%the appendix~\ref{appendix:sNODF_vs_SNES}.}
    \label{fig:bascompte_NODF_vs_SNES}
\end{figure*}

\subsection{Measure differences}
First, in order to study the behaviours of the previous measures, we compare them on the above-mentioned dataset. 
Fig.~\ref{fig:bascompte_NODF_vs_SNES} shows that indeed the normalized SNES is highly correlated with NODF (actually, it is not true for the non normalized version of the spectral nestedness, due to its dependence on the total number of link). In a sense we may think that indeed, while they differ in the philosophy, the two measures are capturing the same structure, as stated in~\cite{Staniczenko2013}. After a detailed comparison with the appropriate null models, we will see that it is not the case. 

\begin{figure*}[htb!]
    \centering
    \includegraphics[width=0.9\textwidth]{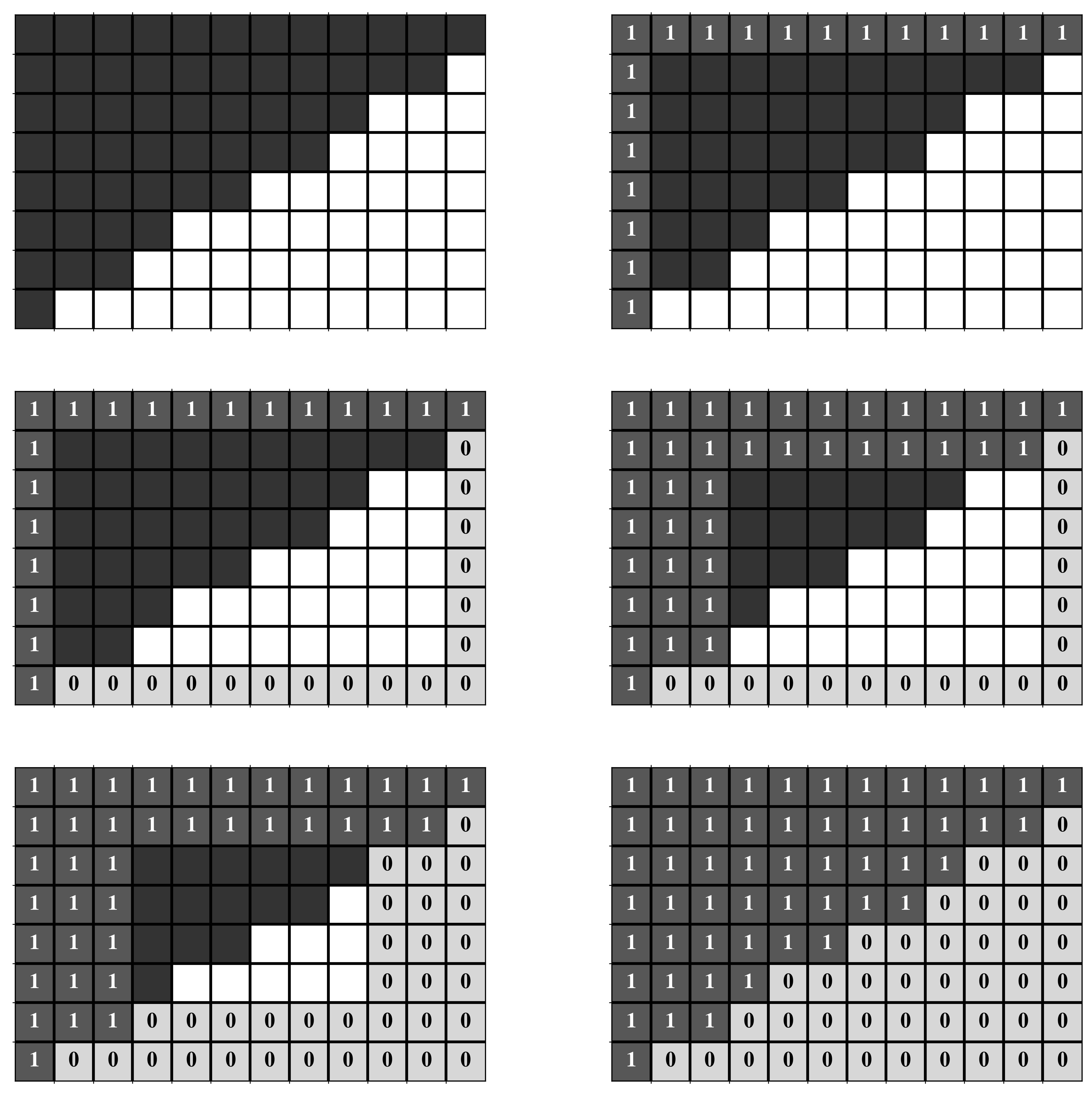}
    \caption{An example of a perfectly nested network with its probabilities per link from the BiCM: at the first step, the first row and column are full, and the degree is respectively 12 and 8. So the link probabilities must be exactly one, for preserving the row sum and the column sum. At the second step, since the last row and column have degree 1, the remaining entries must sum to 0, yielding all zeros. Again, at the third and fourth steps the rows and columns that are completely full or empty univocally determine the respective probabilities to be 1 or 0. At the end of this process, the link probabilities are all set to 0 or 1, so the corresponding canonical ensemble contains only one matrix.}
    \label{fig:PNN}
\end{figure*}

\subsection{Degree sequence vs. nestedness}\label{subsec:degree_sequence}
The degree sequence of the network carries some information about the nestedness of the system, the extreme case being the Perfectly Nested Network (\emph{PNN} in the following) one. Actually, in this case, the degree sequence identifies completely the network and both the micro- and the canonical ensembles are composed by a single network, i.e. the PNN one. This was already observed in~\cite{Lee2016a} for the microcanonical ensemble, but it is surprisingly true also for the canonical ensemble. While the technical details can be found in Section 2 in the Supplementary Information, a pictorial representation is shown in Fig. \ref{fig:PNN}.%\ref{appendix:PNN}
%for the details of this.

Thus in these cases the degree sequence captures the level of nestedness of the whole system, and the statistical significance of the measure loses any value. Actually, even when the network is close to a perfectly nested one, its configuration model ensembles contain a limited number of configurations, that furthermore are all highly nested networks. Thus, a real network may show a high value of the nestedness measure (whatever it is), which is, nevertheless, statistically non significant with respect to a null model discounting the degree sequence: actually in such a case the high value of the nestedness is already captured by the degree sequence. We will examine in more details the role of the null model in the following sections. 

\subsection{Measure and models differences}

\begin{figure*}[ht!]
    \centering
    \includegraphics[width=\textwidth]{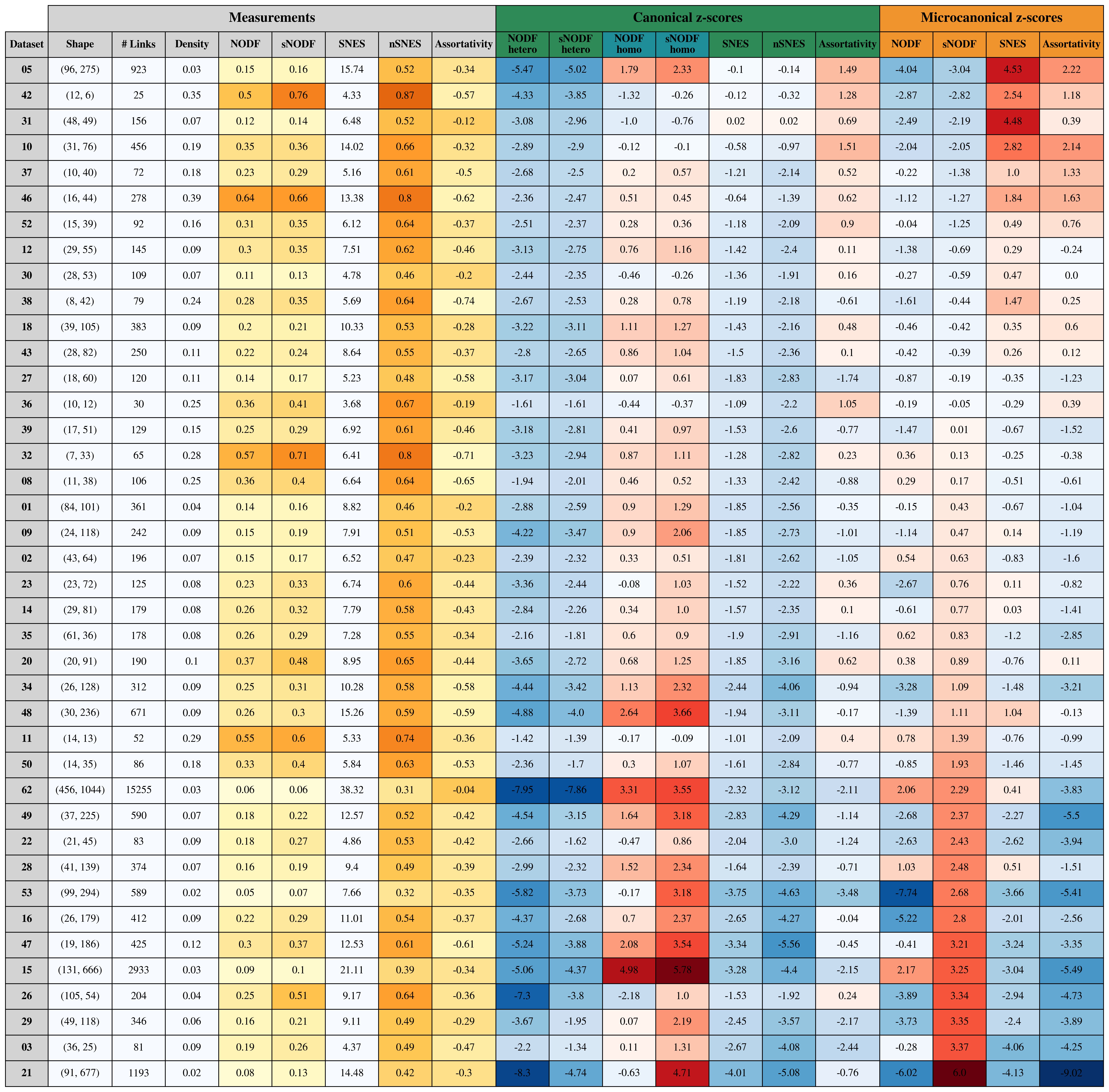}
    \caption{Measures and z-scores of each of the networks in the \textit{Web of Life} pollination binary dataset, ordered according to the sNODF microcanonical z-scores. The microcanonical z-scores of the nSNES are omitted because they are identical to the SNES ones. The color scales have been normalized linearly in the respective measures' domains for the measures, while for the z-scores there is a unique color scale, blue for the negative and red for the positive scores. The SNES measure does not have a color scale since they are not comparable given the different sizes.}
    \label{fig:z_scores_table}
\end{figure*}

\begin{figure*}[ht!]
    \centering
    \includegraphics[width=0.9\textwidth]{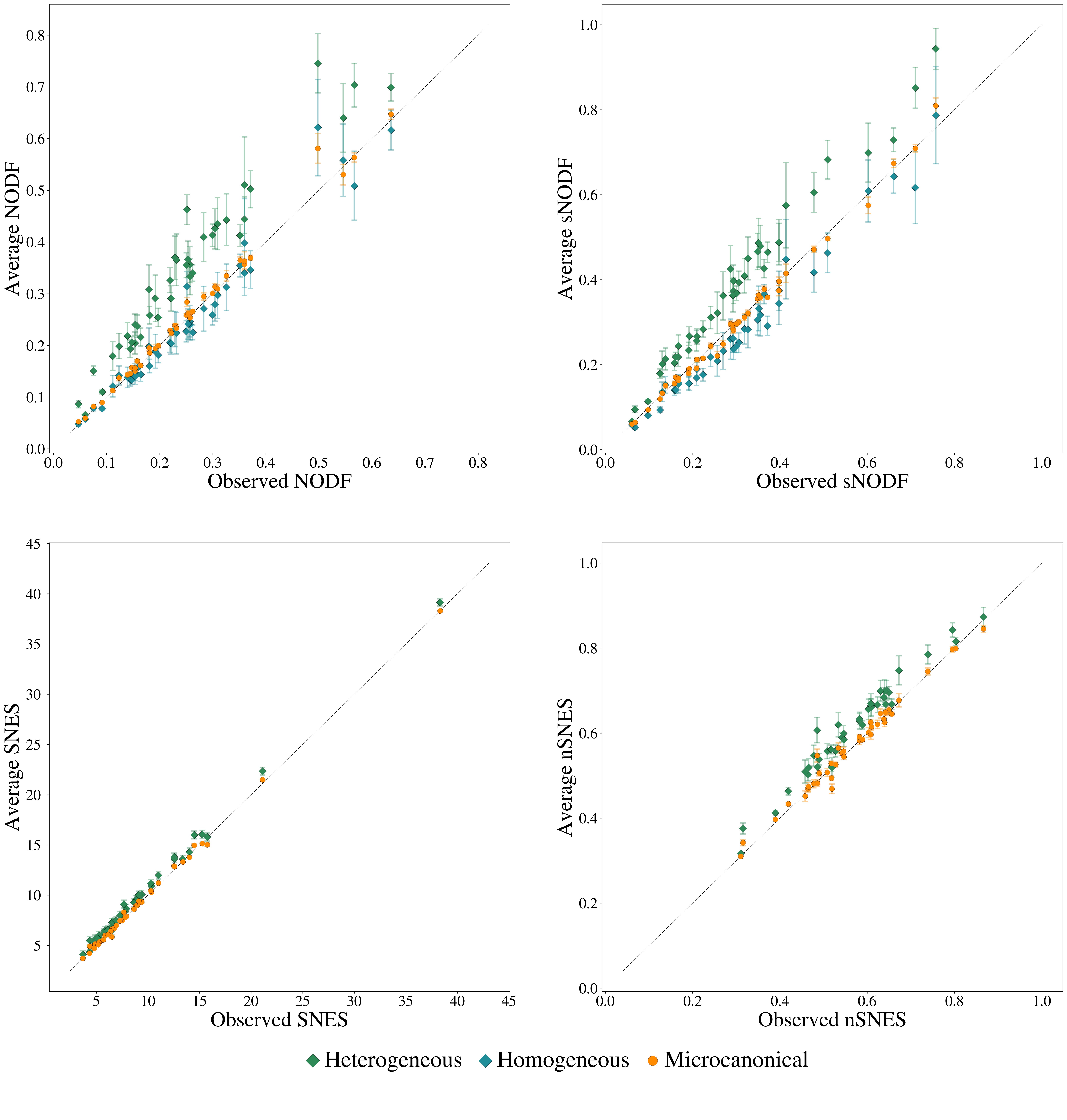}
    \caption{Micro- vs canonical measures for all 40 \textit{Web of Life} pollination datasets: the error bars represent the standard deviations of the respective ensemble.}
    \label{fig:micro_vs_grand}
\end{figure*}

\begin{figure*}[ht!]
    \centering
    \includegraphics[width=0.9\textwidth]{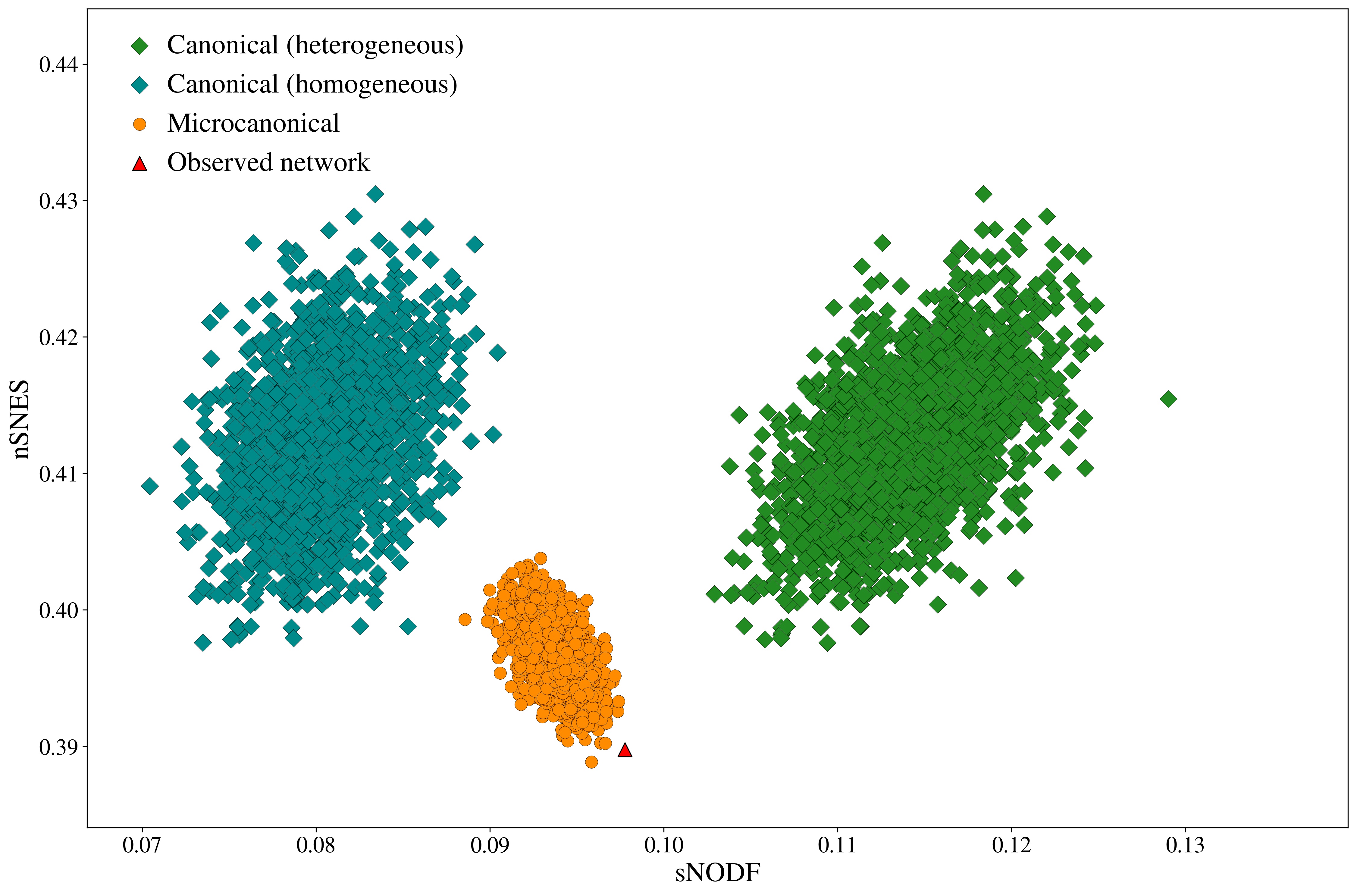}
    \caption{sNODF and nSNES for 2000 realizations of the microcanonical and canonical ensembles, generated from dataset 15\cite{petanidou1993} of the \textit{Web of life} collection. In the two ensembles the measures present opposite correlations, and both the heterogeneous and the homogeneous canonical approaches show a similar correlation between nSNES and sNODF.}
    \label{fig:SNES_NODF_ensemble}
\end{figure*}

\noindent The averages of the measures are systematically different when using a microcanonical model or a canonical one for several reasons. %, since, in the latter case, 
The first observation is that the variance in the degrees of the nodes generates a bias in all quantities that scale non-linearly in the number of links.

Actually, there is another issue generating a bias the canonical ensemble. Indeed, a network sampled from the ensemble can present some isolated nodes that do not contribute to the measurements of both NODF and SNES (and their modifications). Given the steep power law degree distribution of many of the considered networks, this will typically be the case. For further details, in section 3 of the Supplementary Information %~\ref{appendix:isolated}
we analyse the frequency of isolated nodes in generating the canonical ensemble. We will discuss this issue and how it generates a bias in greater detail focusing on each of the two measures in the following paragraphs.\\

\subsubsection{NODF vs. null models}\label{subsubsec:NODF_vs_null_models}
The displacement of the NODF measures between the two ensembles is the result of multiple effects. The most evident bias is caused by the normalization factor that is the denominator in \ref{eq:NODF} and \ref{eq:sNODF}. A network sampled from the BiCM ensemble will have, on average, the same number of links of the original network, but many isolated nodes (see the Supplementary Information Section 3 % \ref{appendix:isolated}
for more details). This is due to the small link probabilities related to nodes of low degree in large networks, that sometimes give rise to an empty row or column in a sampled matrix. These nodes, therefore, do not contribute to the total NODF or sNODF, and one has to choose how to handle the normalization factor in  \ref{eq:NODF} and \ref{eq:sNODF}.\\
If one chooses to consider the number of connected nodes of the sampled network (as in the original definition of the NODF), this will generate a positive bias by having a comparable quantity divided by a lower denominator (the number of the connected nodes in each realisation can be only smaller than the value of the real network). We call this approach \emph{heterogeneous}.\\
Otherwise, considering the normalization factor of the original network will introduce contributions even from isolated nodes, thus altering the philosophy of the original definition. Moreover, such approach will introduce a bias in the opposite direction, dividing by a factor that is larger than what it should be if considering only the connected network. 
We call this approach \emph{homogeneous}.\\ Both choices are equally admissible, depending on the interpretation of the comparisons one wants to follow. Personally, we think that the normalization should not involve the isolated nodes, as in the original definition, i.e. we prefer the heterogeneous normalization. For completeness, in the next subsections we will consider both of them. Interestingly enough their differences do not affect the conclusions.%~\footnote{The authors are grateful to Laura Hernandez, Yamir Moreno and Claudia Payrato-Borras for pointing this out to our attention.}. %We do not consider the effect of disconnected components with more than one node: this can happen in a sampled network, but is typically not very common, and could happen in an ecological system: in fact, only 24 out of 40 networks of our dataset are actually made of a single connected component.}

On top of this, another effect to be considered is the presence of fluctuations in considering the degree sequence. % can be explained by
%the overestimation of the total number of V-motifs by the BiCM.
As mentioned in the previous section, both ensembles contain only one configuration in the case of a perfectly nested degree sequence and their measures are trivially exactly the same. When the two ensembles separate for a non-perfectly nested matrix, the canonical ensemble produces some variance in the degrees of the nodes. %, which generates the effect of \ref{eq:overestimation}. 
This effect is not present in the microcanonical ensemble, where the degrees of all nodes are fixed deterministically.
%The average number of V-motifs~\cite{Diestel2006} scales quadratically in $L$~\cite{Saracco2015} and is therefore actually overestimated in the canonical ensemble. Indeed, the number of observed V-motifs is 
%\begin{equation}
%    N_V^* = \sum_{\alpha}^{N_\Gamma}\binom{h_a^*}{2} = \sum_{\alpha}^{N_\Gamma}\frac{h_\alpha^*(h_\alpha^*-1)}{2}
%\end{equation}
%and thus we have
%\begin{equation}
%\begin{split}
%    \langle N_V \rangle - N_V^* =& \sum_\alpha\frac{\langle h_\alpha^2 - h_\alpha \rangle - \big((h_\alpha^*)^2 - h_\alpha^*\big)}{2}\\ 
%    =& \sum_\alpha\frac{\langle h_\alpha^2 \rangle -(h_\alpha^*)^2}{2}\\ 
%    =& \sum_\alpha\frac{\sigma_{h_\alpha}^2}{2}= \frac{\sigma_L^2}{2}\geq0,
%    \end{split}
%\end{equation}
%and the average total number of V-motifs is systematically overestimated in the canonical ensemble. 
Such an effect has an impact on the NODF and in particular it provides new evidences regarding the non equivalence of the various ensembles.

\subsubsection{SNES vs. null models}\label{par:SNES_overestimation} We observe that the spectral radius is slightly overestimated in the canonical model. %Some first evidences say
Our guess for this behaviour is that on average, out of two matrices with the same number of links, the one with the smallest number of nodes has the largest radius, so when a sample of the canonical model has an empty row or column it has, on average, a higher radius. Some evidences for this behaviour are given in the Supplementary Information Section 4. %~\ref{appendix:SNES_vs_N}.
Still, we are not able to evaluate such discrepancy.
Regarding the nSNES, the overestimation seems to be strongly increased by the normalization factor. Although this is intuitive, since $\langle \sqrt{L} \rangle < \sqrt{\langle L \rangle }$ because of the non-zero variance of $L$, the fact that the spectral radius has an intrinsic dependence on $L$ makes it hard to evaluate precisely.

\subsubsection{The significance of the nestedness measures with respect to the various statistical ensembles}\label{par:sign}
\noindent Bearing in mind all of the considerations of the previous paragraphs, we can interpret the z-scores of the table in Fig.~\ref{fig:z_scores_table}. The four canonical NODF columns of the table refer to the z-scores of the two variants of NODF, with the two different normalizations with respect to the canonical ensemble. A representation of the differences among the models and measures is also given in Fig.~\ref{fig:micro_vs_grand}.\\ 
In the case of the heterogeneous normalization the z-scores are all negative, because of the overestimation of both NODF and sNODF in the ensemble. There are, though, important differences between the NODF and sNODF measures in some cases, which are mainly due to the presence of many nodes with the same degree.\\

For the homogeneous normalization, there is still a certain agreement in the signs of the z-scores of NODF and sNODF, with the same caveat discussed above for the heterogeneous case. In opposition to the heterogeneous columns, the z-scores are positive in most of the networks analysed, in agreement with the discussion of \ref{subsubsec:NODF_vs_null_models}. In this sense, it is striking that the choice of the normalization factor may drive to opposite conclusions, regarding the statistical significance of the measure on the real network.\\
Then we have the columns of SNES: similarly to the heterogeneous normalized NODF, even in this case, the canonical null model has all negative z-scores, due to the slight overestimation of the SNES. %The normalized SNES column also contains all negative z-scores due to its sublinear dependence on $L$, as discussed before. 
The second-last column contains the microcanonical SNES z-scores.\\ 
As it can be observed from the matrix, there is no agreement between the column of the SNES and the sNODF columns in the microcanonical ensemble. Note that for the microcanonical null model, the z-scores of the nSNES were not reported since they correspond to the one of the SNES (the normalizing factor cancels). A hint is given by the assortativity z-scores, whose Pearson correlation coefficient with the  SNES scores is 0.84, while SNES and NODF anti-correlate with a score of -0.88. 

In order to investigate this difference, we generate a scatter plot of the realizations of the different ensembles (Fig. \ref{fig:SNES_NODF_ensemble}), plotting the NODF against the SNES of the sampled networks. The results are striking: the two measures are highly anti-correlated on the microcanonical ensemble, while this effect is hindered by the fluctuations in the canonical ensemble.

\begin{figure*}[ht!]
    \centering
    \includegraphics[width=\textwidth]{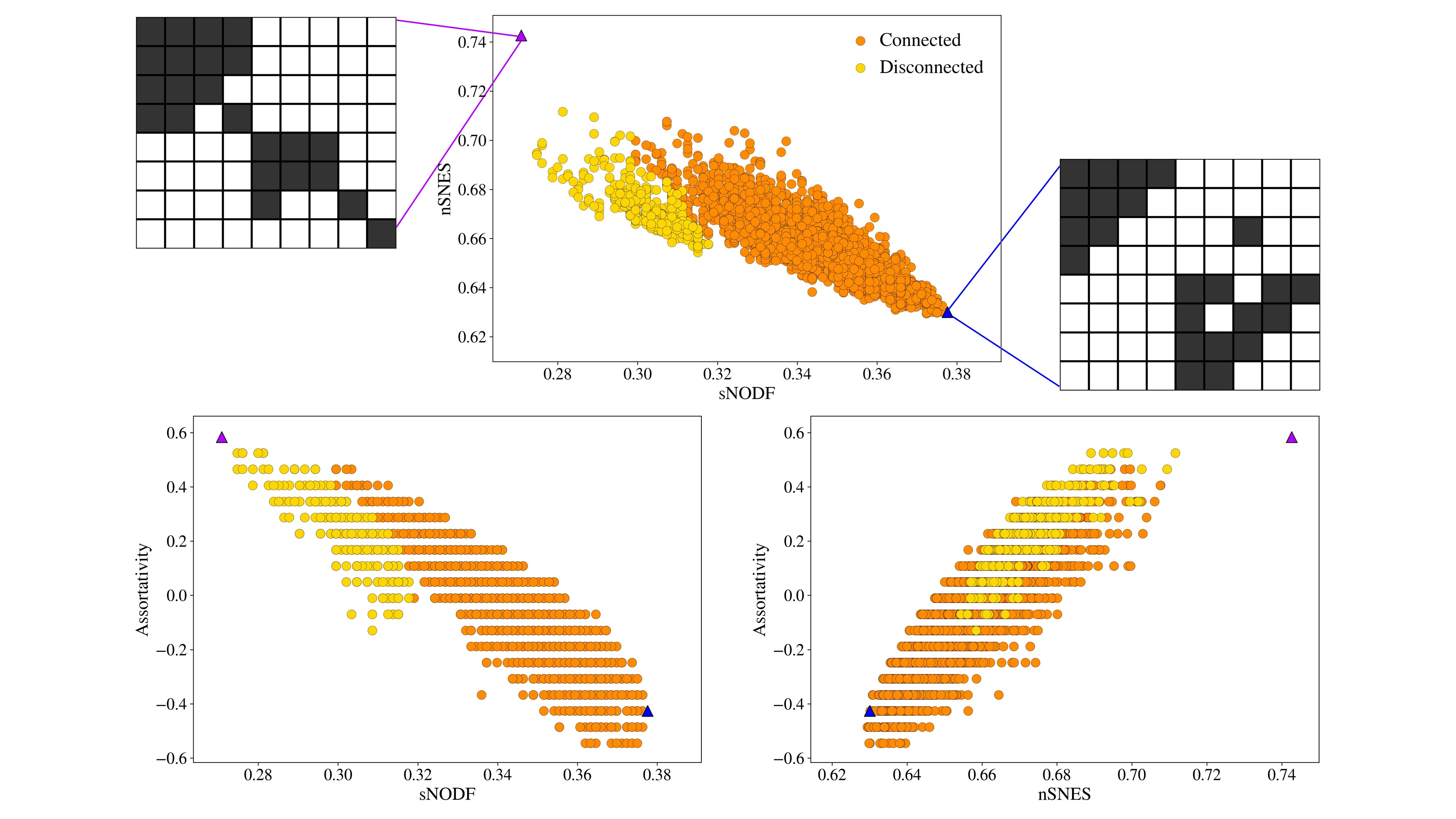}
    \caption{Top: a sample from the microcanonical configuration model ensemble with the relative scores of nSNES and sNODF. Bottom: the same sample with scores of sNODF against assortativity (left), nSNES against assortativity (right). Different colors are used for sampled networks that result connected or disconnected. The highlighted networks have high values of nSNES or sNODF, and show that for their extreme values, the systems can be disconnected (left) or barely connected (right). The left matrix, with a high nSNES, has a really assortative configuration, while the right one has a high sNODF and is highly disassortative. We do not exclude disconnected networks in our analysis since it could be a possible configuration for an ecological system.}
    \label{fig:NODF_SNES_synthetic_micro}
\end{figure*}

Using the other proposed measures, the results are always similar when comparing a NODF measure and a spectral nestedness measure. NODF and SNES are actually capturing different ways of being ``nested". This is easily seen on a synthetic very small network, of size $8\times 9$. We generate a sample from the microcanonical model and see how the matrices maximizing NODF and SNES are made (Fig. \ref{fig:NODF_SNES_synthetic_micro}).

\begin{figure*}[htb!]
    \centering
    \includegraphics[width=\textwidth]{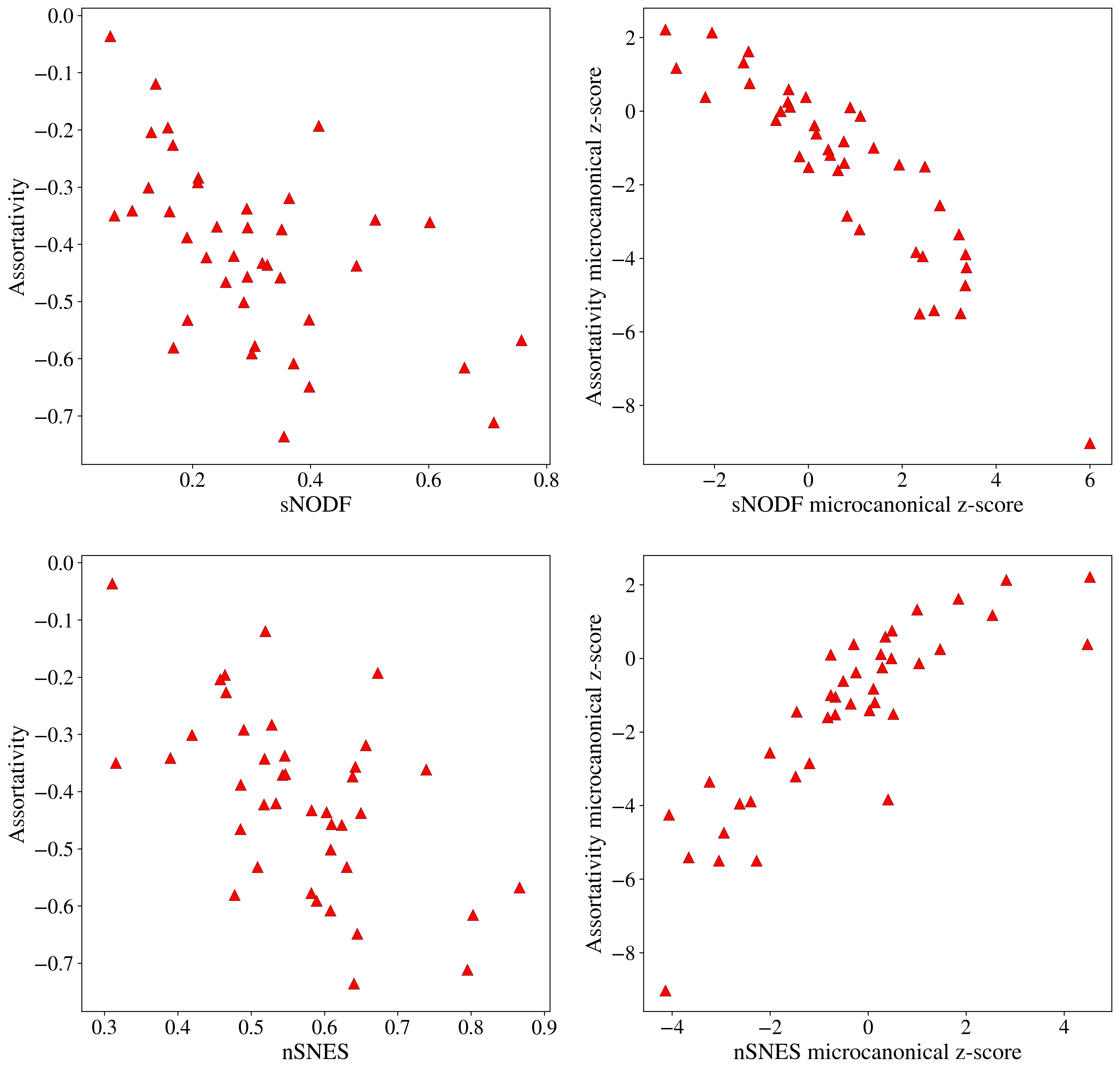}
    \caption{sNODF and nSNES vs assortativity. The correlation between assortativity z-scores and microcanonical nestedness z-scores is not well captured by the raw measures. The Spearman correlation coefficients are: top left -0.56, top right -0.89, bottom left -0.49, bottom right 0.86.}
    \label{fig:sNODF_vs_assortativity_zscore}
\end{figure*}

Actually the NODF-maximizing matrix has one of the smallest value of assortativity, while the one that maximizes the SNES presents a big hub of the highest degree nodes and two smaller disconnected subgraphs, see Fig.~\ref{fig:NODF_SNES_synthetic_micro}. Roughly speaking, on the one hand, the SNES prefers networks in which highly connected nodes link to highly connected nodes, since they are sort of carrying the ``mass" of the adjacency matrix (which is what the spectral radius is measuring). On the other hand the NODF, due to the denominator of its contributions, prefers to link poorly connected nodes with highly connected ones, thus focusing on disassortative configurations. Regarding the anti-correlation between the NODF (or similar definition) and the assortativity, other studies got to similar conclusions~\cite{Jonhson2013,Abramson2011}; as far as we know, there were no evidences regarding the opposite behaviour of the SNES.\\ 
Let us underline that the (anti)correlation between the NODF or SNES and the assortativity is present only when discounting microcanonically the degree sequence: in real data, such correlation is not evident, as Fig.~\ref{fig:sNODF_vs_assortativity_zscore} shows. Otherwise stated, the (anti)correlation is present only when the contribution of the degree sequence is discounted. More details regarding the correlation between assortativity and the various nestedness metrics can be found in the Supplementary Information Section 5: Fig. D in the Supplementary Information shows examples for other networks in the \textit{Web of Life} dataset.%~\ref{appendix:assortativity_vs_nestedness}.

\section{Discussion}\label{sec:Discussion}
While the abstract idea of nestedness in networks is quite straightforward, %its mathematical definition is less trivial. 
we saw that a mathematical definition capturing the degree of nestedness of a real system is much less trivial.
As a consequence, while nested structures are ubiquitously observed across several networks, measuring the actual level of nestedness along with its statistical significance remains a challenging task.\\
In the present manuscript we investigated in details different metrics of nestedness in both real-world and synthetic networks. In particular, we mainly focused on two measures, NODF~\cite{Almeida-Neto2008a} and SNES~\cite{Staniczenko2013}, and some of their modifications~\cite{MARIANI20191}. When applied to real networks, these metrics go in the same direction, as they give positively correlated results.\\

We then moved to discount the contribution of the degree sequence to the different nestedness measures. Literally, according to the case of study and to the available information on our system we can create suitably chosen series of randomized copies of our graph (ensembles). This procedure allows us to use the machinery of Statistical Physics to assess the significance of our measurements.\\
Thus, for our aim,  we can define null models preserving the degrees of nodes either as hard (microcanonical ensembles~\cite{Strona2014}) or as soft (canonical ensembles~\cite{SquartiniTiziano2017,Cimini2018}) constraints.
Otherwise stated, we are using the extensions of the microcanonical and canonical ensembles to complex networks in order to discount the information carried by the node degrees: the degree sequence is supposed to have an effect on the nestedness~\cite{Jonhson2013,payrato2019breaking}, thus we want to focus on the information carried by the different metrics that cannot be explained by the degree sequence only. Let us remark that the null models implemented are \emph{ergodic}, i.e. they explore the phase space uniformly.\\   

First, we concentrated our attention on Perfectly Nested Networks (PNN). A PNN has a degree sequence that admits only a single network, i.e. the PNN itself, irrespective of whether the degrees are treated as hard or soft constraints: both the microcanonical and canonical ensembles of a PNN are composed by the PNN network only. Otherwise stated, there exist perfectly nested degree sequences and each of them defines univocally a single network, i.e. the PNN one. 
In the case of PNNs, thus, the value of the nestedness is completely due to the degree sequence only. But what happens when the network is not perfectly nested?\\

We compared the values of NODF and SNES measured on real networks with the expectations of, respectively, the microcanonical and canonical ensembles. 
As theoretically demonstrated in other studies~\cite{DenHollander2000,Barre2007,Campa2009,Radin2013,Touchette2015,Squartini2015c, Squartini2015a}, the two ensembles are not equivalent, thus they should be characterized by different macroscopic properties. 
Literally, we found that the two families of nestedness metrics are negatively correlated when the microcanonical ensemble is used, while they are positively correlated in the canonical ensemble. Actually, the fluctuations of the canonical ensemble cover the real behaviours of NODF and SNES. Instead, once the degree sequence is fixed as a hard constraint, the level of nestedness is influenced by higher-order correlations between the degrees themselves, and in particular the assortativity of the network. Indeed, the two classes of measures of nestedness give different results in the microcanonical ensemble, when considering networks with different assortativity: NODF tends to give larger values of nestedness when the network is disassortative, while SNES tends to give larger values of nestedness when the network is assortative. Otherwise stated, we present an example in which the same information, i.e. the degree sequence, is \emph{ergodically} discounted, but the sign of the measured correlations are opposite if compared with a microcanonical of with a canonical null model.\\

Thus, 
other than the choice of the measure, if checking for the statistical significance of the nestedness of a system, one should make a principled choice of the ensemble used as a null model in the analysis~\cite{Squartini2015c, Squartini2015a}. 
The microcanonical ensemble, which treats degrees as hard constraints, should be preferred if the observed degrees are error-free, i.e. if they are the actual values of the property to be kept fixed in the null hypothesis. If one suspects that the observed degrees are instead subject to some sort of error (e.g. measurement errors, incomplete data collection, poor sampling, etc.), then the microcanonical ensemble should be avoided, as it will give zero probability to the true (undistorted) configuration and to any configuration with the same degree sequence as the true configuration. 
%Therefore, if the observed degrees are possibly `noisy', one should prefer the canonical ensemble, which treats degrees as soft constraints and can access the true configuration with nonzero probability (as desirable, this probability is higher for lower levels of `noise', i.e. for a lower distortion in the values of the degrees). 
In this case, we suggest to use the measures that present the smallest biases for fluctuating degrees, i.e. the SNES and the homogeneous sNODF or NODF.\\

Let us finally remark that our paper does not provide any indication on which is the nestedness metric that should be used, or on which is the right null model to be implemented in order to state the statistical significance of the nestedness measured. In a sense, each nestedness measure has contraindications, and every null model, even if discounting the same information, has its peculiar properties. In this sense, it is crucial to know exactly the behaviour of the ingredients we are handling. Our contribution is in highlighting odd behaviours, previously undetected, that, if not under control, can take to unjustified conclusions.

Nevertheless, in light of our results, the question remains on how to tackle the problem of the nestedness in real system, even after a proper and justified choice of the nestedness measure. An easy solution can be, once the null model has been chosen, to report for the chosen nestedness measure, both the average over the ensemble and the z-scores on the real network. The former value provides an evaluation of the nestedness as encoded by the degree sequence, the latter how significant is the observed nestedness, once the degree sequence is discounted.

\section{Acknowledgements}
GC and FS acknowledge support from the European Project SoBigData++ GA. 871042 and the TOFFEe PAI (Progetto di Attività Integrata) project funded by the IMT School Of Advanced Studies Lucca. GC also acknowledges support from European Project Humane-AI-net 952026. DG acknowledges support from the Dutch Econophysics Foundation (Stichting Econophysics, Leiden, the Netherlands) and the Netherlands Organisation for Scientific Research (NWO/OCW). CJT acknowledges financial support from University of Zurich through the URPP Social Networks. The authors are thankful to Laura Hernández, Yamir Moreno and Claudia Payrató-Borràs for discussions and useful suggestions.

\section{Data availability}
This work has used the Web of Life dataset (\textit{www.web-of-life.es}).

\section{Author contributions}
M.B. performed the analysis, all authors designed and interpreted the analysis and wrote the manuscript.

\section{Additional information}
The authors declare no competing interests.

\section{Correspondence}
All correspondence should be sent to M.B.

\bibliography{Bibliography.bib}

\newpage
% \thispagestyle{empty}
% \mbox{}
% \newpage

{\large\textbf{Supplementary Information}}

\renewcommand\thefigure{\Alph{figure}}
\addtocounter{figure}{-\value{figure}}

\renewcommand\thesection{\arabic{section}}
\addtocounter{section}{-\value{section}}

\renewcommand\theequation{\Roman{equation}}
\addtocounter{equation}{-\value{equation}}

\thispagestyle{empty}

\section{sNODF vs. SNES}\label{appendix:sNODF_vs_SNES}
\begin{figure*}[htb!]
    \centering
    \includegraphics[width=\textwidth]{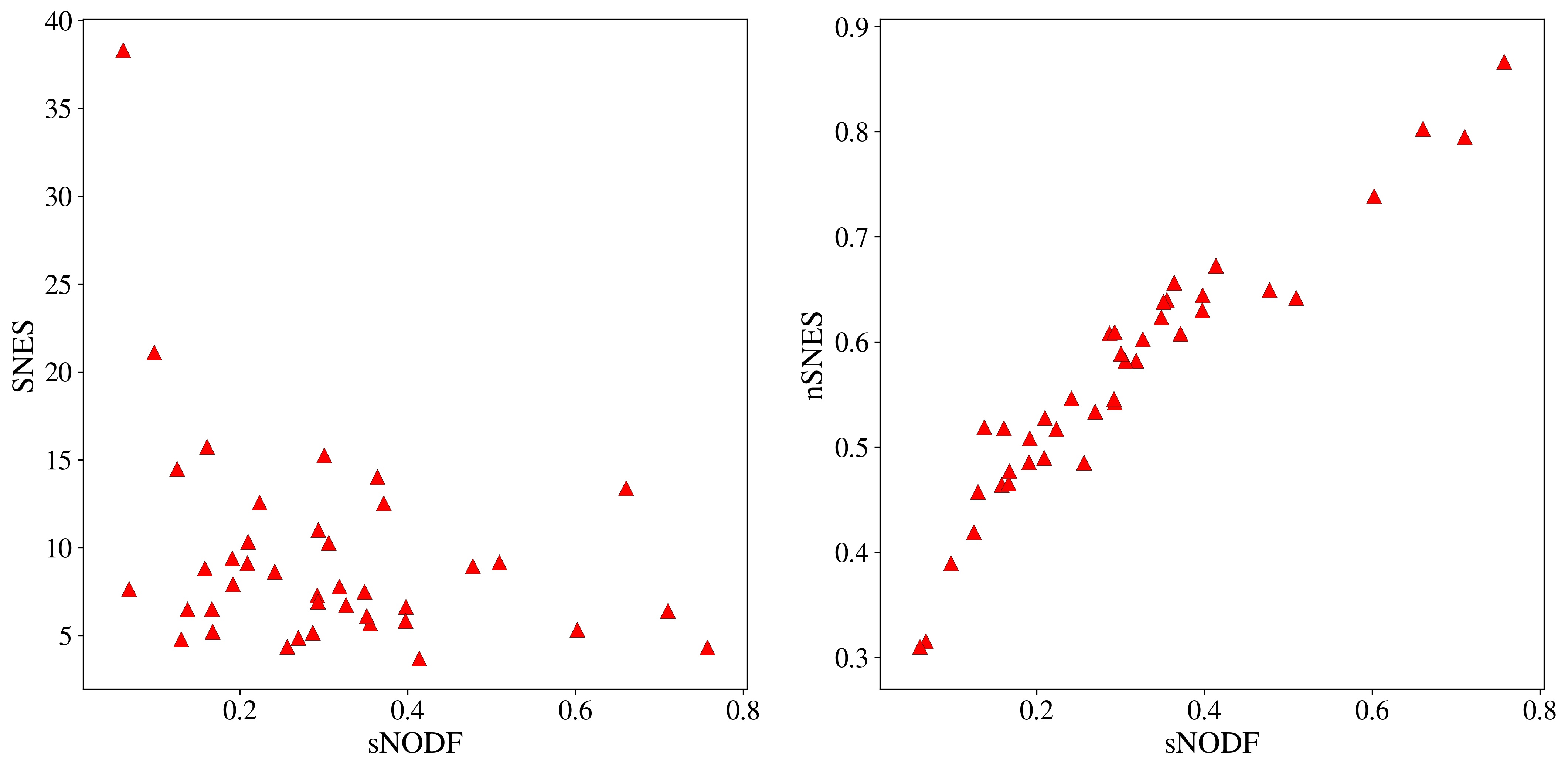}
    \caption{sNODF vs SNES (left) and vs nSNES (right) for the 40 networks of the Bascompte dataset. Spearman correlation coefficients are, respectively, -0.26 and 0.96. The figure is very similar to Fig. 1 %\ref{fig:bascompte_NODF_vs_SNES}
    in the main text, so we included only one in it.}
    \label{fig:bascompte_sNODF_vs_SNES}
\end{figure*}
In the main text we showed the correlation between the NODF and the SNES, in its two different normalizations. In Fig.~\ref{fig:bascompte_sNODF_vs_SNES} it is possible to observe that an analogous relation is present between the sNODF and the SNES measures.

\section{Perfectly Nested Networks}\label{appendix:PNN}
For a PNN, the corresponding ensembles (both microcanonical and canonical) are singular, i.e. the only matrix in the ensembles is the PNN itself. In this section we explain this fact.\\
Let us start from the microcanonical ensemble. 
First, let us summarise the main steps of the Curveball:
\begin{enumerate}
    \item Select at random a couple of nodes on the same layer (for making the example clearer let us consider, in full generality,  $i,\,j\in\NL$);
    \item Check that the neighbourhoods of the nodes are not perfectly overlapping: if so, start again.
    \item Take the set of uncommon neighbours $U(i,j) =\{\alpha\in N_\Gamma|(m_{i\alpha}=0\,\&\&\, m_{j\alpha}=1)||(m_{i\alpha}=1\,\&\&\, m_{j\alpha}=0)\}$ and remove them from the neighbourhood of both;
    \item Assign $k_i - \sum_\alpha m_{i\alpha}m_{j\alpha}$ new neighbours to node $i$, chosen at random from $U(i, j)$ and the rest of the nodes in $U(i, j)$ to node $j$.
\end{enumerate}

Consider the case in which $k_i=k_j$: due to PNN nature, $U(i,j)=\emptyset$ and the algorithm stops at the step 2. Then, consider the case $k_i>k_j$: $U(i,j)$ contains only the connections that $i$ has and $j$ has not (due to the perfect nestedness of the network, all connections of $j$ are connections of $i$ too). Then, at step 4, the number of new neighbours of $j$ is $k_j-\sum_\alpha m_{i\alpha}m_{j\alpha}=0$, while the same quantity is exactly $|U(i,j)|$ for $i$, thus the algorithm is stuck in the present configuration. A similar intuition can be found in~\cite{Lee2016a}.\\

In the canonical ensemble the situation is a little more involved. Let us consider, as an example, the biadjacency matrix in Fig. 2 %~\ref{fig:PNN}
in the main text, representing a PNN; the presented arguments can be generalised to any PNN. Due to the ordering we imposed on the biadjacency, if rows and columns represent respectively the L and the $\Gamma$ layers, we have:
\begin{equation}
\begin{split}
\langle k_1\rangle=&\sum_\alpha^{N_\Gamma} p_{1\alpha}=k_1^*=N_\Gamma;\\
\langle h_1\rangle=&\sum_i^{\NL} p_{i1}=h_1^*=\NL,
\end{split}
\end{equation}
which can be satisfied if and only if $p_{1\alpha}=1,\,\forall\alpha\in\Gamma$ and $p_{i1}=1,\,\forall i\in\text{L}$. Thus all entries involving the fully connected nodes are deterministic. Such a conclusion has implications, on the opposite side of the biadjacency matrix:
\begin{equation}
\begin{split}
\langle k_{\NL}\rangle=&\sum_{\alpha>1}^{N_\Gamma} p_{\NL\alpha}+1=k_{\NL}^*=1;\\
\langle h_{N_\Gamma}\rangle=&\sum_{i>1}^{\NL} p_{iN_\Gamma}+1=h_{N_\Gamma}^*=1,
\end{split}
\end{equation}
which, in turns, implies $p_{\NL\alpha}=0,\,\forall\alpha>1\in\Gamma$ and $p_{i N_\Gamma}=0,\,\forall i>1\in\text{L}$, i.e. the entries of nodes with only a single connection are deterministic too. Then let us pass to consider again the first nodes:
\begin{align}
\langle k_2\rangle=&1+\sum_{\alpha>1}^{N_\Gamma-1} p_{2\alpha}+0=k_2^*=N_\Gamma-1;\label{eq:k_2}\\
\langle h_2\rangle=\langle h_3\rangle=&1+\sum_{i>1}^{\NL-1} p_{i2}+0\nonumber\\
=&1+\sum_{i>1}^{\NL-1} p_{i3}+0\label{eq:h_2}\\
=&h_2^*=h_3^*=\NL-1\nonumber
\end{align}
(in the second line we use the fact that columns 2 and 3 have the same degree, thus their Lagrangian multipliers are equal and so $p_{i2}=p_{i3},\,\forall i\in\text{L}$). Let us first focus on equation \ref{eq:k_2}: we have $N_\Gamma-2$ unknown probabilities, summing to $N_\Gamma-2$. Thus $p_{2\alpha}=1$ for $1<\alpha<N_\Gamma$. Analogous considerations are valid for all $p_{i2}$s and $p_{i3}$s and thus these entries are again deterministic. Iteratively discounting the information obtained at the previous steps, it is possible to show that the canonical ensemble of a PNN is composed by a single graph, or, more correctly, the probability for every representative in the ensemble is 0 but for the PNN itself (which, instead has $P(PNN)=1$).\\

\section{Isolated nodes in the canonical ensemble}\label{appendix:isolated}
\begin{figure*}[hb!]
    \centering
    \includegraphics[width=\textwidth]{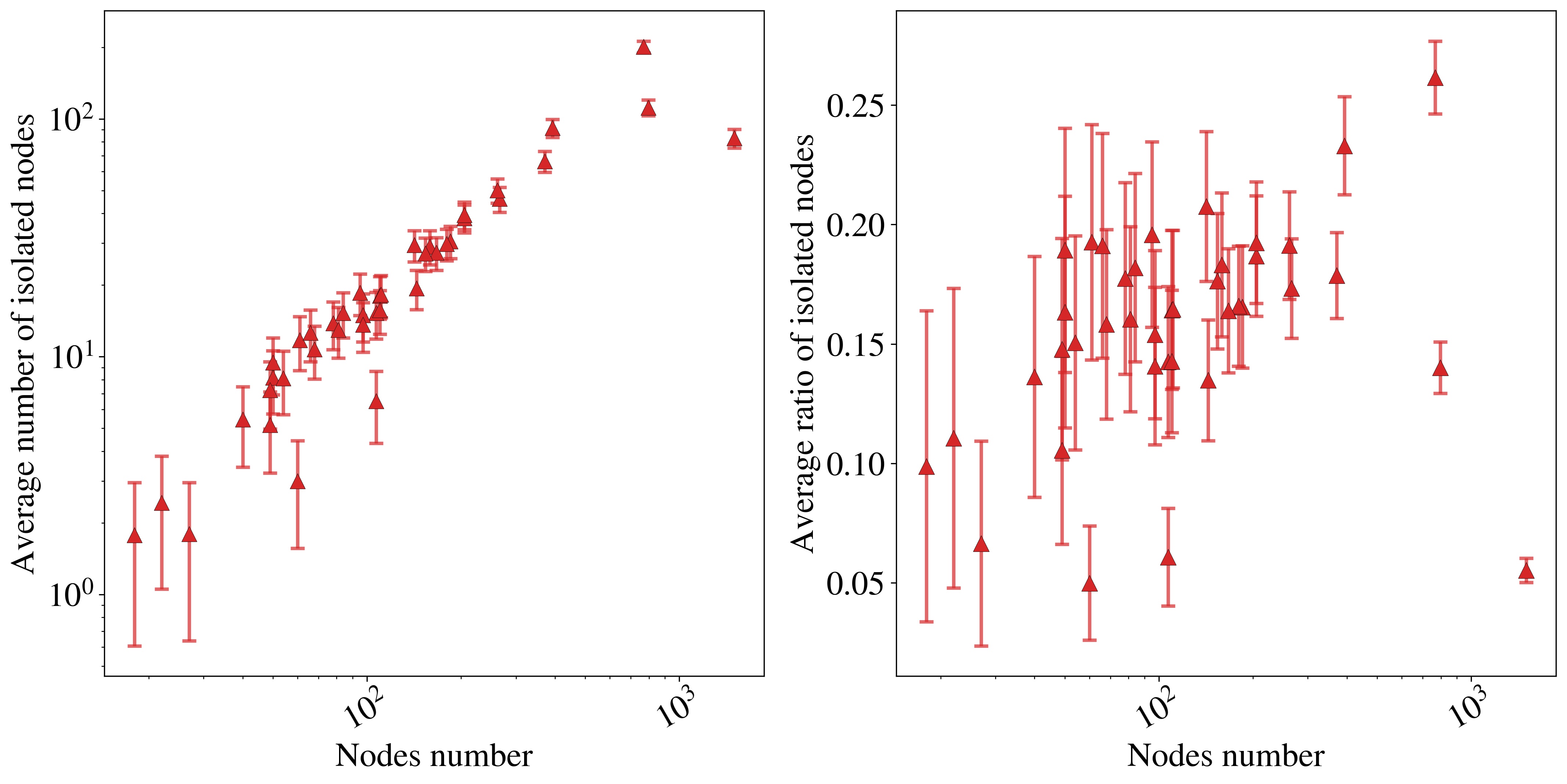}
    \caption{The average number (left) and ratio (right) of isolated nodes in the canonical ensemble samples as a function of the total nodes, with error bars representing one standard deviation. The relative Spearman correlation coefficients are 0.96 (left) and 0.36 (right).}
    \label{fig:missing_nodes}
\end{figure*}

In the canonical ensemble, the degree sequence is fixed on average, thus there are fluctuations from realisation to realisation. Therefore, nodes with low degree, say close to 1, in the real network, can result as isolated in some realisations of the ensemble. As it can be seen from Fig.~\ref{fig:missing_nodes}, the average of isolated nodes over the size of the network is nearly constant all over the dataset.

\section{SNES dependence on the number of nodes}\label{appendix:SNES_vs_N}
\begin{figure}[htb!]
    \centering
    \includegraphics[width=.5\textwidth]{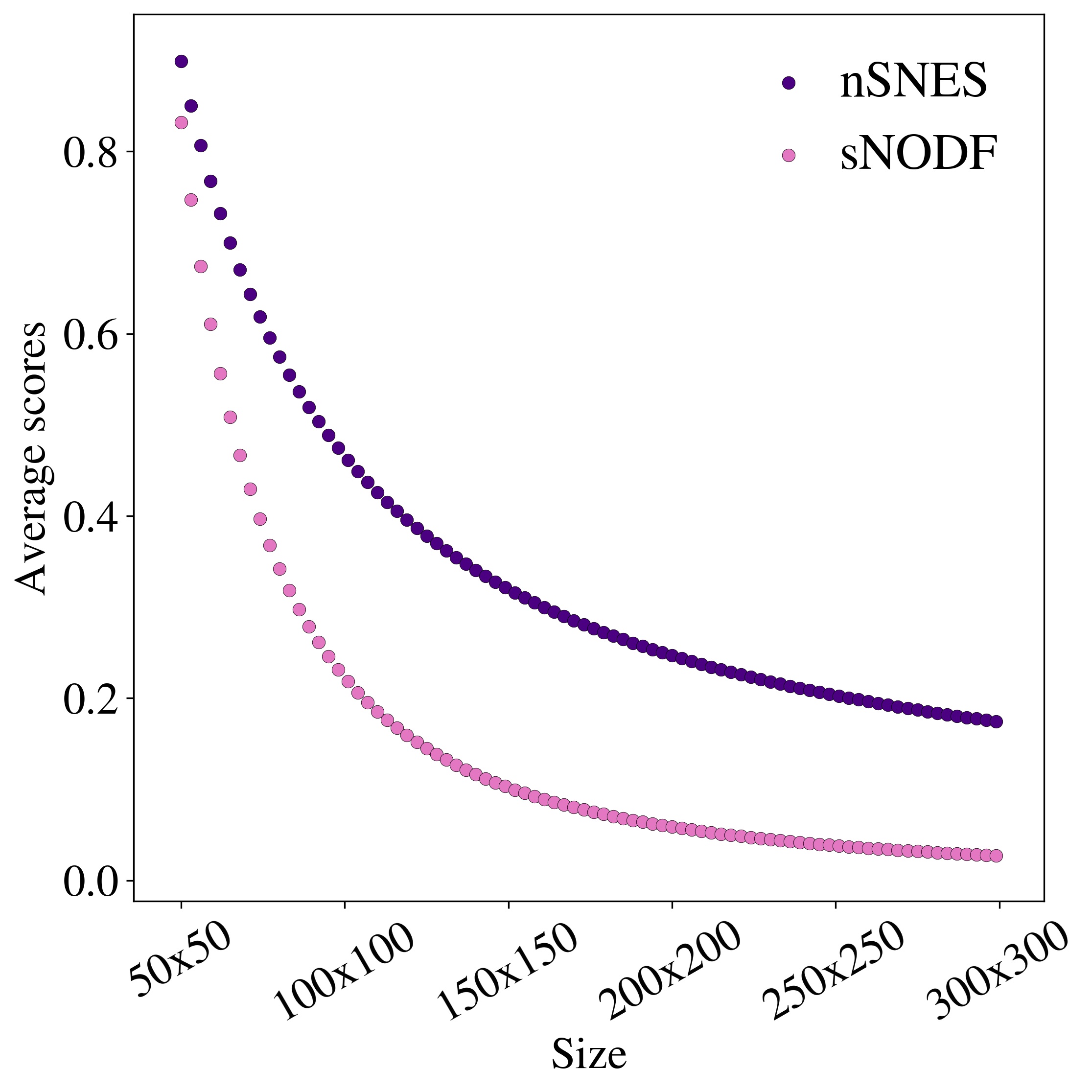}
    \caption{In this experiment we generate random bipartite networks of various sizes, filling them with exactly 2000 links in random positions. For every size considered, 1000 samples have been generated and we measured the resulting average nSNES and sNODF. We omit the corresponding standard deviations because they are negligible. Although this is not a rigorous argument since many of the considered networks could have isolated nodes, it indicates that reduced average dimension with fixed average links number can generate a bias in the nSNES and NODF/sNODF measures in the canonical ensemble.}
    \label{fig:decreasing_measures}
\end{figure}

In the paragraph III.C.2  %~\ref{par:SNES_overestimation}
in the main text, we observed on the canonical samples that spectral radius is a little greater than the true value. %Some first evidences say
Our intuition is that on average, out of two matrices with the same number of links, the one with the smallest number of nodes has the largest radius. Fig. \ref{fig:decreasing_measures} shows a little experiment confirming our guess: we generate synthetic networks of various sizes, but with the same number of links. As it can be seen, as the size increases, the average nSNES reduces. Something similar happens for the sNODF.

\section{Canonical vs microcanonical ensemble: more examples of ensembles distributions}\label{appendix:canonical_vs_microcanonical}

\begin{figure*}
    \centering
    \includegraphics[width=\textwidth]{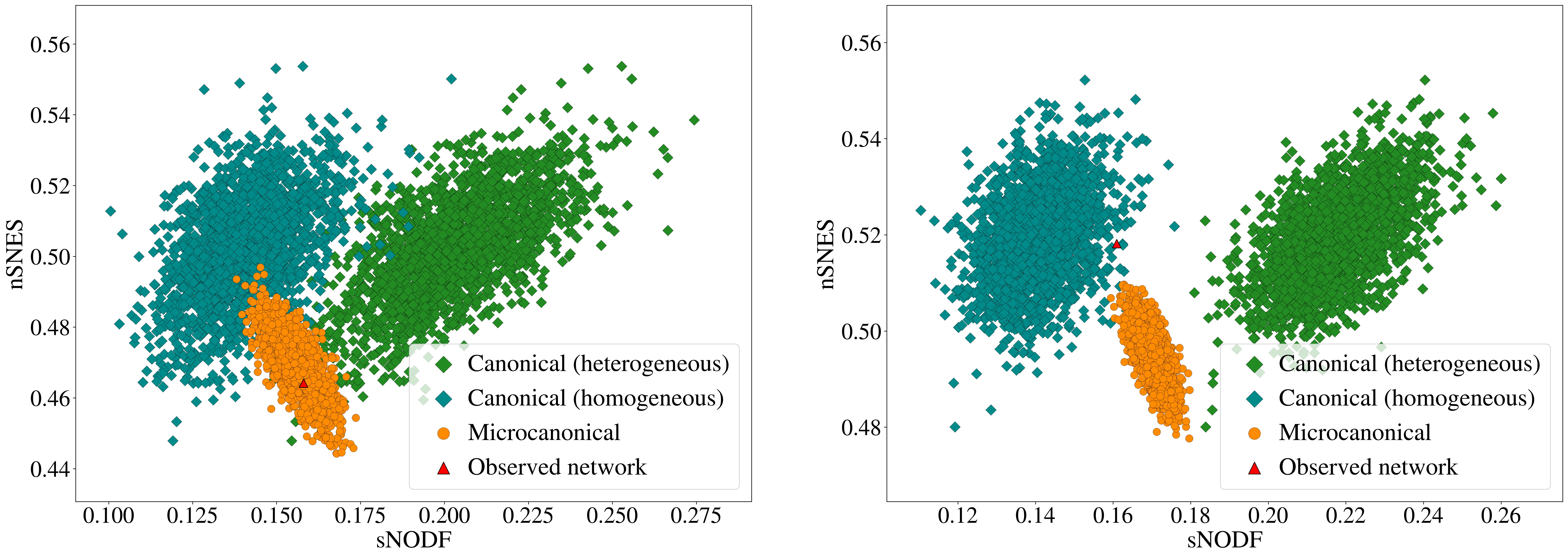}
    \caption{The equivalent of Fig. 5 of the main text but for two more networks, specifically dataset 5 (left) and 1 (right). The position of the real network values with respect to the ensembles may vary.}
    \label{fig:SNES_NODF_ensemble_examples}
\end{figure*}

In Fig. 5 in the main text we showed the realizations of the canonical and microcanonical ensembles. In Fig. \ref{fig:SNES_NODF_ensemble_examples} we show the same plot for two more real networks. While the observed values are not as extreme as the one presented in the Fig. 5 of the main text, still the same behaviours are present: a positive correlation between the nSNES and sNODF in the canonical ensemble and a negative one in the microcanonical case. Let us  remark that the observed values of both nSNES and (heterogeneous) sNODF represent the negative extremes in the case of the canonical ensemble in the dataset 5 (left panel of Fig. \ref{fig:SNES_NODF_ensemble_examples}), while they fall almost in the center of the distribution of the microcanonical ensemble.  Even more striking is the case of the dataset 1 (right panel of Fig. \ref{fig:SNES_NODF_ensemble_examples}): while the observed (heterogeneous) sNODF is extremely low both for the canonical and microcanonical ensembles, the observed nSNES is much greater than expected in the microcanonical ensemble, while it stays in the middle of the distribution in the canonical one. As already mentioned in the main text, the value of the homogeneous and heterogeneous sNODF have opposite behaviours in the canonical ensemble, as it can be seen in both panels of Fig. \ref{fig:SNES_NODF_ensemble_examples}.

% {\color{blue}\section {Assortativity  vs nestedness in the canonical ensemble}
% In the Fig.~\ref{fig:nodf_snes_canonical} we present the analogous of Fig.~6 in the main text for the canonical ensemble. As it can be seen from the second and third panel, almost no correlation between the assortativity and the various nestedness measures can be observed, while the correlation between the two nestedness metrics is evident in the first panel. Also in this case, the overestimation of the measures in the canonical ensemble screens the anticorrelation observed in Fig.~6.}

\begin{figure*}
    \centering
    \includegraphics[width=\textwidth]{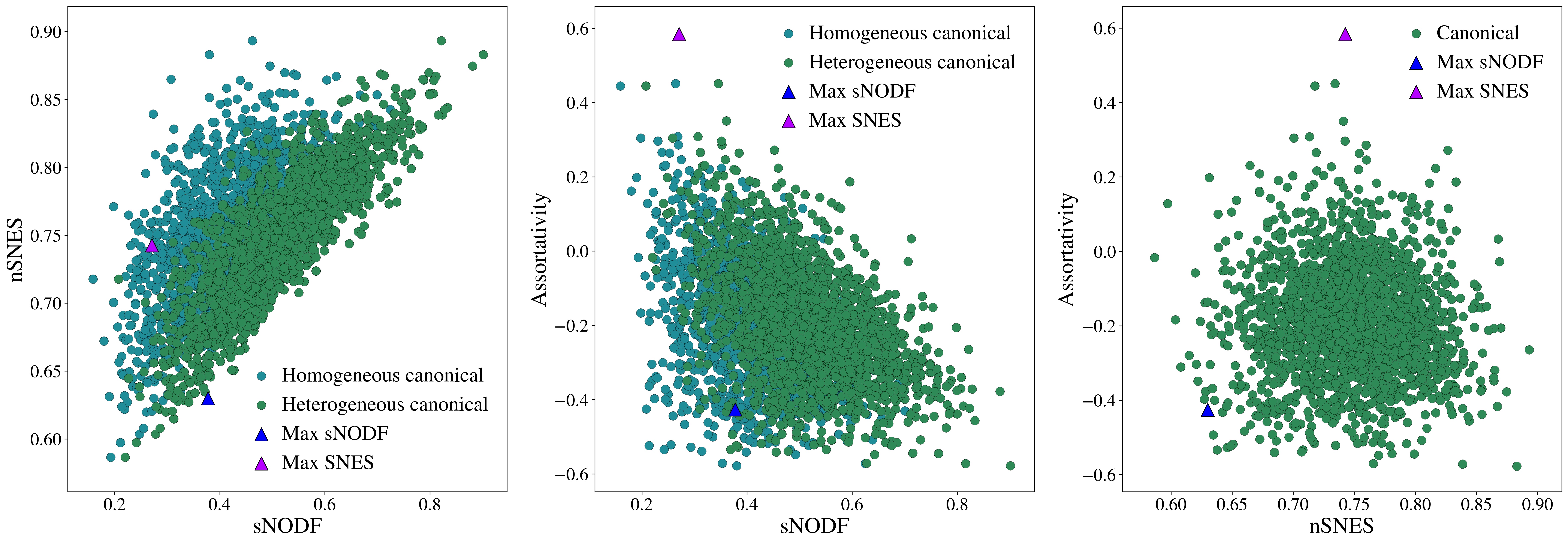}
    \caption{The equivalent of Fig. 6 of the main text but for the canonical ensemble. The blue and violet realizations are the matrices of maximum sNODF and SNES of the microcanonical ensemble.}
    \label{fig:nodf_snes_canonical}
\end{figure*}

%\newpage
\section{Assortativity vs. nestedness in the microcanonical and canonical ensemble}\label{appendix:assortativity_vs_nestedness}

In the paragraph III.C.3 %~\ref{par:sign}
in the main text, we observed a high correlation between nSNES and the assortativity of the sampled networks, and an anticorrelation of both with the sNODF in the microcanonical ensemble. In Fig. 7 % \ref{fig:sNODF_vs_assortativity_zscore}
of the main text, in the panels on the left, we show that this is not as evident on the measurements of our dataset without filtering and both nSNES and sNODF show a weak anticorrelation with the assortativity.
When, instead, the measures on the real data are compared with the microcanonical ensemble, a strong correlation is evident. In particular, the z-scores of the sNODF (SNES) anticorrelate (correlate) with the z-scores of assortativity, showing different behaviours for the two metrics, see the two panels on the right of Fig. 7 of the main text.%~\ref{fig:sNODF_vs_assortativity_zscore}.

\begin{center}
\begin{figure*}[htb!]
    \centering
    \includegraphics[width=\textwidth]{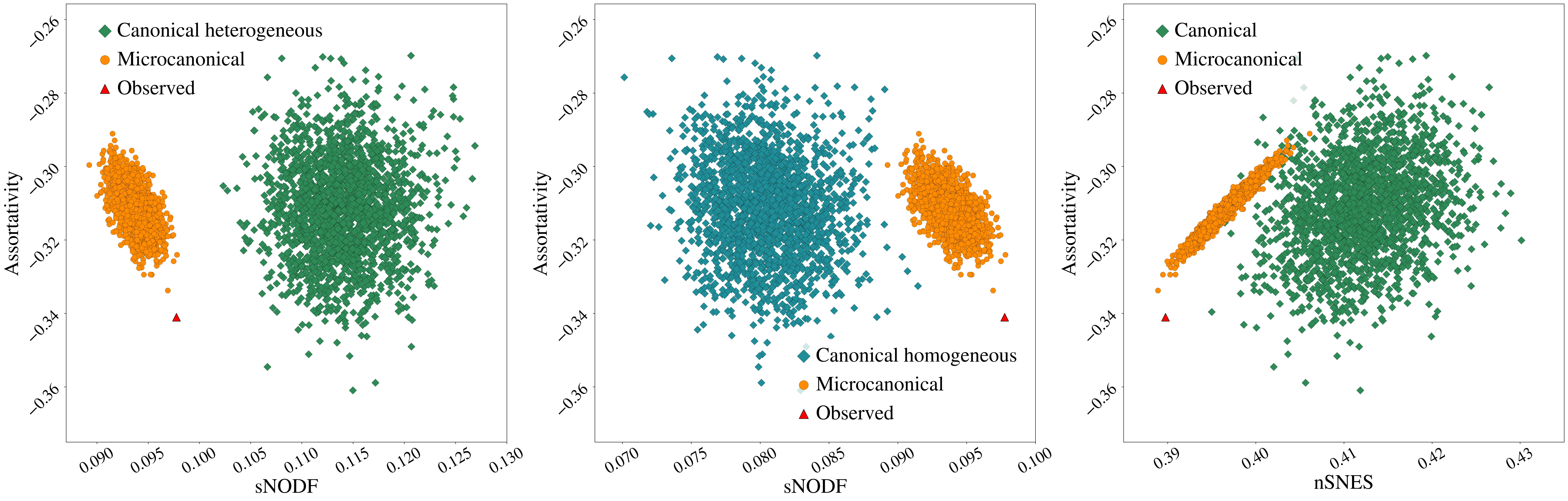}
    \caption{The correlation between the nestedness measures and assortativity is hidden in the canonical ensemble. The Spearman correlation coefficients are, from top to bottom: -0.51 and -0.04 for the top figure, -0.17 and -0.51 for the middle figure, 0.97 and 0.23 for the bottom one.}
    \label{fig:assortativity_correlations}
\end{figure*}
\end{center}

In the canonical ensemble instead, the correlations that we find in the microcanonical ensemble are almost completely lost, as it can be seen in Fig. \ref{fig:assortativity_correlations}. Here the fluctuations cover completely the relations between the assortativity and the various nestedness metrics.

Another example of the behaviour of assortativity in the canonical ensemble is provided in Fig.~\ref{fig:nodf_snes_canonical}, in which we present the analogous of Fig.~6 in the main text for the canonical ensemble. As it can be seen from the second and third panels, almost no correlation between the assortativity and the various nestedness measures can be observed, while the correlation between the two nestedness metrics is evident in the first panel. Also in this case, the overestimation of the measures in the canonical ensemble screens the anticorrelation observed in Fig.~6.

\section{Homogeneous vs microcanonical sNODF}
\begin{center}
\begin{figure*}[htb!]
    \centering
    \includegraphics[width=\textwidth]{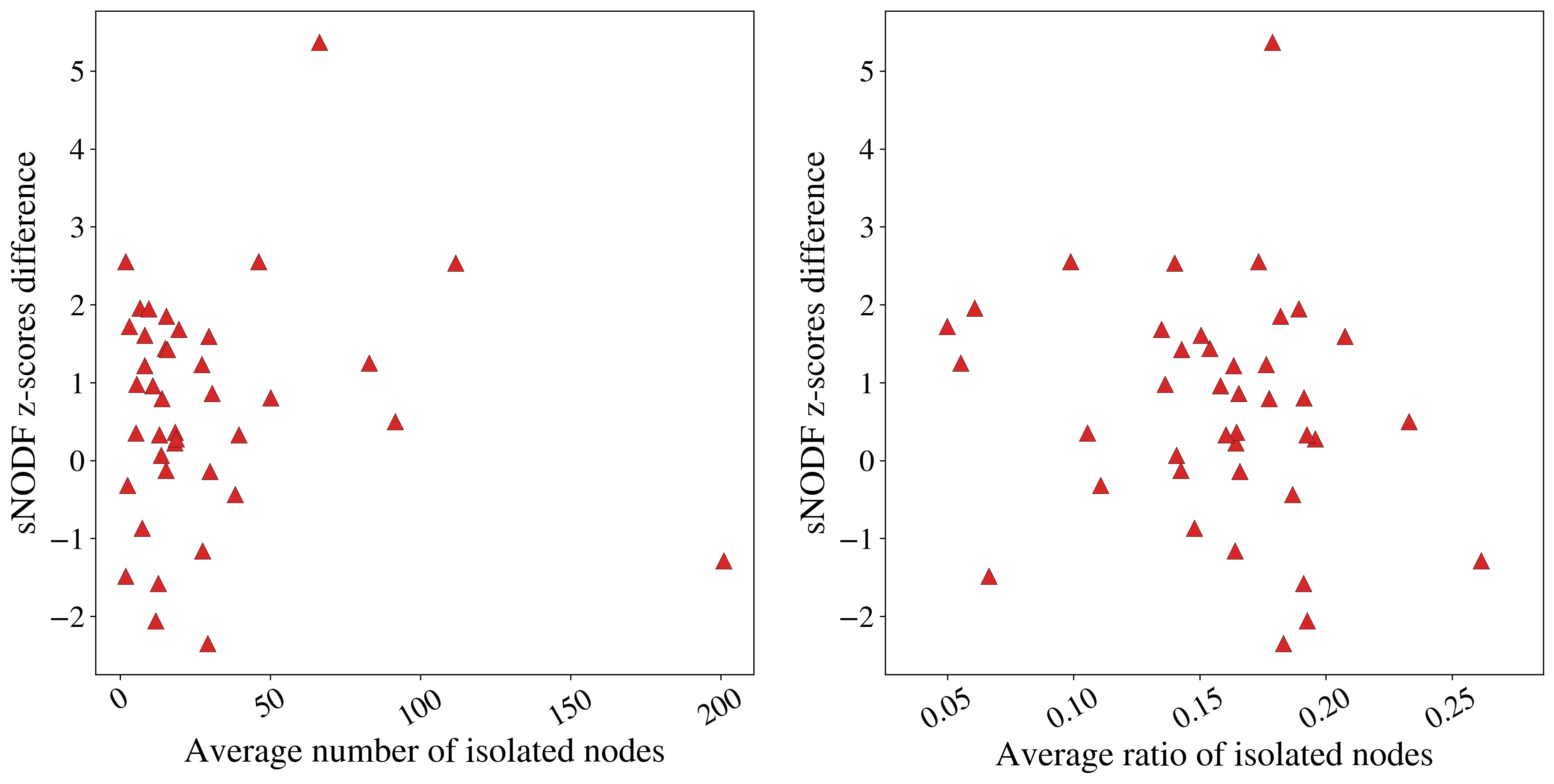}
    \caption{The correlation between the number of isolated nodes in the samplings and the difference between the sNODF homogeneous canonical z-scores and the sNODF microcanonical z-scores. Neither the number of isolated nodes nor their ratio seem to be factors, with Spearman correlation coefficients of 0.02 and -0.25 respectively%, while the assortativity z-score seems to correlate with it, with a coefficient of 0.59
    .}
    \label{fig:z_scores_diff}
\end{figure*}
\end{center}

Although the homogeneous sNODF seems correlated with the microcanonical one by looking at Fig. 3 in the main text, quantifying their differences is not an easy task. In Fig. \ref{fig:z_scores_diff} we show that the difference in the respective z-scores seems to be uncorrelated with the average number of isolated nodes in a network sampled from the canonical ensemble.

\newpage

\end{document}